\documentclass[floats,aps,epsf,amsfonts]{revtex4}
\usepackage[dvips]{graphicx}
\begin{document}

\title{Precession effect of the gravitational self-force in a Schwarzschild spacetime and the effective one-body formalism}
\author{Leor Barack$^1$, Thibault Damour$^2$ and Norichika Sago$^3$}
\affiliation{
$^1$School of Mathematics, University of Southampton, Southampton
SO17 1BJ, United Kingdom, \\
$^2$Institut des Hautes \'Etudes Scientifiques, 35, route de Chartres,
                91440 Bures-sur-Yvette, France \\
$^3$Yukawa Institute for Theoretical Physics, Kyoto University,
Kyoto 606-8502, Japan}

\date{\today}

\begin{abstract}
Using a recently presented numerical code for calculating the Lorenz-gauge gravitational self-force (GSF), we compute the $O(m)$ conservative correction to the precession rate of the small-eccentricity orbits of a particle of mass $m$ moving around a Schwarzschild black hole of mass ${\mathsf M}\gg m$. Specifically, we study the gauge-invariant function $\rho(x)$, where $\rho$ is
defined as the $O(m)$ part of the dimensionless ratio $(\hat\Omega_r/\hat\Omega_{\varphi})^2$ between the squares of the radial and azimuthal frequencies of the orbit, and where  $x=[Gc^{-3}({\mathsf M}+m)\hat\Omega_{\varphi}]^{2/3}$ is a gauge-invariant measure of the dimensionless gravitational potential (mass over radius) associated with the mean circular orbit. Our GSF computation of the function $\rho(x)$ in the interval $0<x\leq 1/6$ determines, for the first time, the {\em strong-field behavior} of a combination of two of the basic functions entering the Effective One Body (EOB) description of the conservative dynamics of binary systems. We show that our results
agree well in the weak-field regime (small $x$) with the 3rd post-Newtonian (PN) expansion of the EOB results, and that this agreement is improved when taking into account the analytic values of some of the logarithmic-running terms occurring at higher PN orders. Furthermore, we demonstrate that GSF data give access to higher-order PN terms of $\rho(x)$ and can be used to set useful new constraints on the values of yet-undetermined EOB parameters. Most significantly, we observe that an 
{\em excellent global representation} of $\rho(x)$ can be obtained using a simple `two-point' Pad\'{e} approximant which combines 3PN knowledge at $x=0$ with GSF information at a single strong-field point (say, $x=1/6$).  
\end{abstract}

\maketitle


\section{Introduction}

The problem of calculating the gravitational self-force (GSF) acting on a pointlike particle 
of mass $m$, in  bound orbit around a Schwarzschild black hole (of a much larger mass ${\mathsf M}$), is now well understood at leading order in the small mass ratio $q \equiv m/{\mathsf M}$ (see \cite{Barack:2009ux,Detweiler:2009ah} for reviews of recent developments). Explicit numerical computations of the GSF, specialized to circular orbits, were carried out by at least three different groups using a variety of methods \cite{BS,Detweiler:2008ft,Keidl:2010pm}. These calculations all rely on mode-sum regularization \cite{Barack:1999wf,Barack:2001gx} but are otherwise based on very different computational strategies which are formulated in different gauges. Two of us recently reported a numerical code for computing the GSF along generic (bound) geodesics with arbitrary eccentricities \cite{Barack:2010tm}. 

In principle, knowledge of the GSF (along with the metric perturbation associated with the particle) gives a complete information about the $O(q)$ post-geodesic dynamics of the binary system. However, the GSF itself is a gauge-dependent notion \cite{Barack:2001ph}, and it is generally not a straightforward task to translate the raw GSF data (i.e., the value of the various components of the GSF along the orbit, computed in a given gauge) into physically meaningful gauge-invariant statements about the orbital dynamics. It is particularly non-trivial to obtain such a gauge-invariant description for the {\em conservative} part of the dynamics, which is also (in principle) accessible from the GSF. For circular orbits, Detweiler  \cite{Detweiler:2008ft} proposed to measure the conservative effect of the GSF using the gauge-invariant relation between the ``red-shift'' parameter $u^t$ (the Schwarzschild-$t$ contravariant component of the four-velocity along the circular geodesic) and the azimuthal frequency $\hat\Omega_{\varphi}$. Using numerical data for the GSF in the Regge-Wheeler gauge, Detweiler was able to compute the relation $u^t(\hat\Omega_{\varphi})$ through $O(q)$. Numerical computations of the GSF in the Lorenz-gauge and in the radiation gauge were later shown to reproduce the same relation $u^t(\hat\Omega_\varphi)$ \cite{SBD,Keidl2}. 

The recent development of a GSF code for eccentric orbits (formulated in the Lorenz gauge) enabled the computation of additional gauge invariant quantities associated with the conservative effect of the GSF. By considering slightly-eccentric orbits, two of the authors have computed the $O(q)$ conservative shift in the orbital frequency of the innermost stable circular orbit (ISCO) \cite{ISCOLetter,Barack:2010tm}. This quantity (unlike $u^t$) has a clear physical interpretation, and, at least in principle, a measurable effect on the signature of gravitational waves from the binary system. Most recently, the GSF correction to the precession rate of orbits with arbitrary eccentricities has also been calculated \cite{BSprep}.

Analyzing the gauge-invariant effects of the GSF allows one not only to compare between GSF calculations carried out in different gauges, but also to make an important connection with the well-established post-Newtonian (PN) theory. Indeed, in Ref.\ \cite{Detweiler:2008ft} Detweiler used the relation  $u^t(\hat\Omega_\varphi)$ to demonstrate, for the first time, that conservative GSF results (for circular orbits) were consistent with the analytic predictions of PN theory. This comparison was further improved, and pushed to higher PN orders, in Refs.\ \cite{Blanchet:2009sd} and \cite{Blanchet:2010zd}.

Such ``cross-cultural'' comparisons (to use the language of Ref.\ \cite{Blanchet:2009sd}) are motivated in several ways. (i) They provide highly non-trivial opportunities to confirm some of
the basic aspects of both GSF and PN formalisms (such as the regularization procedures underpinning them). (ii) Comparison of concrete results from GSF and PN computations serves to test both sets of results; such two-way validation tests are crucial given the complexity of both GSF numerical codes and PN analytic computations. (iii) Results from GSF calculations can inform the development of accurate Analytical Relativity (AR) models by providing high-accuracy calibration data for yet-unknown high-order PN terms. Combined also with Numerical Relativity (NR) results for comparable-mass binaries, one eventually hopes to develop a reliable 
and accurate AR model across the full range of binary mass ratios (e.g., for gravitational-wave detection applications). Within such a program, GSF calculations provide a crucial ``data point'' at the extreme edge (extreme mass ratio, small separation) of the essential parameter space of the binary problem.
(iv) Since PN theory is formulated for arbitrary mass ratios, it could be used to test 
second-(and higher-)order GSF computations when these become available in the future.  

A fundamental challenge in any GSF--PN comparison arises from the fact that the two methods are designed to explore distinct regimes of the binary problem: GSF methods work best for strong-field orbits (the necessary computational resources tend to increase fast with the orbital radius \cite{BS}), while the standard PN series is essentially an asymptotic expansion about the limit of infinite binary separation---although it performs surprisingly well at rather small separations, it ultimately fails to converge at sufficiently small radii.  The upshot is that, in order to facilitate a useful GSF--PN synergy, one must (generally speaking) stretch GSF techniques beyond their natural domain of applicability (i.e., obtain large-radius data) while at the same time attempt to compute new higher-order terms in the PN series, at least through $O(q)$. That this can be achieved in practice was indeed demonstrated by Blanchet {\it et al.}\ in Refs.\ \cite{Blanchet:2009sd,Blanchet:2010zd}.

Although PN theory is the basis of most of our current analytical knowledge of the motion and
radiation of gravitationally interacting comparable-mass binary systems, it is unable, by itself, to provide an analytical description of the entire evolution of such systems. 
Indeed, the PN description is only accurate during the early inspiral and it breaks down during the late inspiral, well before the merger. The Effective One Body (EOB) formalism \cite{Buonanno:1998gg,Buonanno:2000ef,Damour:2000we,Damour:2001tu}
was proposed as a flexible AR framework for describing
the motion and radiation of coalescing binaries over the entire merger process, from the early inspiral, right across the innermost stable orbit, through the `plunge' and to the merger and 
final ringdown. 

At the heart of the EOB formulation is an ``effective Hamiltonian'' which depends on 3 initially-unspecified functions: $A(u;\nu)$, $\bar D(u;\nu)$ and $Q(u,p_r;\nu)$
 (see Ref.\ \cite{Damour:2009ic} for a review). Here $u\equiv G({\mathsf M}+m)/(c^2 r_{\rm EOB})$ is the dimensionless gravitational potential (with $r_{\rm EOB}$ being the EOB measure of the binary separation),
$\nu\equiv m {\mathsf M}/(m + {\mathsf M})^2$ is the symmetric mass ratio, and $p_r$ is the EOB (relative) radial momentum. PN theory gives access to the expansions
 of these functions in powers of the variable $u$. These ``Taylor-expanded'' functions
 were found to behave badly (at the 3PN level) in the strong-field region $u \sim 1/6$ 
 \cite{Damour:2000we}. It was then suggested to replace them with suitably {\em resummed}
 expressions (Pad\'e approximants). The natural analytical flexibility of the EOB
 formalism leads one to consider resummed functions, especially $A(u;\nu)$, which include yet uncalculated coefficients, $a_5, a_6, ...$ parametrizing PN terms beyond the currently known 3PN level. These a priori unknown coefficients can then be ``calibrated'' by comparing EOB predictions to numerical data from NR simulations \cite{Damour:2002qh,Buonanno:2007pf,Damour:2009kr,Buonanno:2009qa}.
 Vigorous ongoing activity in this area is convincingly demonstrating  the utility of the EOB framework to describe accurately all phases of the merger process of comparable-mass binaries,
  from the early inspiral to the final ringdown. Moreover, EOB theory has been shown to perform very well also in the extreme mass ratio regime \cite{Damour:2007xr,Yunes:2009ef,Bernuzzi:2010ty}.

The recent GSF results provide a new source of calibration data for the EOB model. The GSF data have the potential to be particularly useful for this purpose, for a number of reasons. (i) Unlike most available NR data, GSF data  is ``clean'' and very accurate---the numerical component in GSF calculations only involves {\em linear} differential equations. (ii) In GSF calculations, again unlike in NR, it is straightforward to disentangle the conservative aspects of the dynamics from the dissipative ones (EOB treats these two aspects separately). (iii) GSF data ``fill the gap'' in the strong-field/extreme-mass-ratio corner of the parameter space, which (as of yet) is inaccessible to either PN or NR. (iv) GSF calculations already return accurate data for orbits with large eccentricities; this data could give a good handle on the two EOB functions ($\bar D$ and $Q$) which describe the radial component of the motion and have not been studied in detail so far. On the other hand, the GSF data have the shortcoming of only giving access to the $O(\nu)$ terms in the $\nu$-expansions of the EOB functions $A(u;\nu)$, $\bar D(u;\nu)$ and $Q(u,p_r;\nu)$.

The prospect for a useful GSF--EOB synergy was recently highlighted  in a paper by one of us \cite{Damour:2009sm}. This work explored how the single GSF data point obtained in Ref.\ \cite{ISCOLetter} [namely, the $O(q)$ conservative shift in the ISCO frequency] constrains
the shape around $u=1/6$ of the $O(\nu)$ term in the crucial EOB radial potential $A(u;\nu)$,
and thereby can improve the determination of two higher-order parameters  $a_5$ and $a_6$, which previous EOB--NR analysis had found to be strongly degenerate. Ref.\ \cite{Damour:2009sm} also identified several additional gauge-invariant quantities that might be computable using current GSF technology, and could provide useful additional calibration data for EOB.  

In this work we consider one of the gauge-invariant quantities proposed in Ref.\ \cite{Damour:2009sm}, namely the {\em conservative GSF correction to the periastron precession for slightly eccentric orbits}. More precisely, we consider the related quantity $\rho(x)$, defined as the $O(q)$ conservative part of the squared ratio of the radial and azimuthal frequencies [see Eq. (\ref{Wexp}) below], with $x$ being the dimensionless gravitational potential defined from the invariant azimuthal frequency. The function $\rho(x)$ is a truly dynamical gauge-invariant characteristic of the conservative dynamics, in that,
as shown in \cite{Damour:2009sm}, its knowledge gives a direct handle on the strong-field behavior of [the $O(\nu)$ part of] a combination of the EOB functions $A(u;\nu)$ and $\bar D(u;\nu)$ . The GSF computation of the function $\rho(x)$ in the interval $0<x\leq 1/6$, which we present here, is therefore the first determination of the {\em strong-field behavior} of a combination of functions which are of
direct significance for describing the conservative dynamics of binary systems \footnote{Refs.\ \cite{Detweiler:2008ft,Blanchet:2009sd,Blanchet:2010zd} did determine the strong-field behavior
of the gauge-invariant function $u^t(\hat\Omega_\varphi)$, but this function cannot, as far as we know,
be used to inform the AR description of the dynamics of binary systems.}. As such it can contribute to the development of accurate AR models of the gravitational wave signals emitted
by coalescing binaries.

The {\it weak-field} behavior of $\rho(x)$, on the other hand, has already been analyzed \cite{Damour:1999cr,Damour:2009sm} through 3PN order [$\rho^{\rm PN}(x)=\rho_2 x^2+\rho_3 x^3 + O(x^4 \ln x)$]. In addition, the logarithmic contributions present at the 4PN and 5PN levels
($\rho_4^{\rm log} x^4 \, \ln x+ \cdots+ \rho_5^{\rm log}  x^5 \, \ln x + \cdots$) have been
recently determined analytically \cite{DamourLogs}. [The presence of logarithmic contributions
in the (near-zone) PN expansions, starting at the 4PN level, follows from old work \cite{Anderson:1982fk,Blanchet:1985sp,Blanchet:1987wq}. The
importance and detectability of these logarithmic terms in the comparison between GSF data 
and PN expansions has been recently discussed in Refs. \cite{Damour:2009sm,Blanchet:2010zd}.]
Here we shall use the numerical methods of Refs.\ \cite{ISCOLetter,Barack:2010tm} to obtain corresponding GSF data for $\rho(x)$,
both in the strong-field region ($x \sim 1/6$) and in the weak field ($x \ll 1$),
and explore what can be learned by comparing the GSF results with the EOB/PN expressions. 

The paper is structured as follows. In the next section we derive an expression for the
function $\rho(x)$ in terms of GSF quantities, and review the EOB expression for $\rho(x)$. In Sec.\ III we describe the numerical method applied to obtain the necessary GSF data, and present our numerical results for $\rho(x)$. In Sec.\ IV we test the numerical data against the currently available EOB/PN results, and in Sec.\ V we examine to what extent the GSF data can be used to determine yet-unknown higher-order PN terms in the EOB model. Section VI explores the utility of simple Pad\'{e} models (i.e., rational-function fits), based on a minimal amount of EOB and GSF information, to provide accurate global fits for $\rho(x)$. Section VII contains a summary and a discussion of future directions for EOB--GSF synergy.

Throughout this paper we use standard geometrized units (with $G=c=1$), metric signature ${-}{+}{+}{+}$, and Schwarzschild coordinates $(t,r,\theta,\varphi)$.

\section{Small-eccentricity precession effect in the GSF and EOB Formulations}

We consider a gravitationally-bound binary comprising a particle of mass $m$ and a Schwarzschild black hole of a much greater mass, ${\mathsf M}\gg m$. We denote the small mass ratio by $q\equiv m/{\mathsf M}(\ll 1)$, and throughout our analysis work through $O(q)$ only. Readers should be wary of the conflict between the standard PN/EOB notation and the common GSF one (see Table \ref{table:masses}); our notation here represents a compromise between the two sets of notations. Note that the symmetric mass ratio \footnote{In some of the PN/EOB literature the
symmetric mass ratio is denoted $\eta$.} is $\nu \equiv q/(1+q)^2 = q + O(q^2)$, so that an
$O(q)$ quantity is also $O(\nu)$. One should, however, beware of the important $O(q)$ difference 
between the large mass ${\mathsf M}$ (which is often used in GSF works to adimensionalize
frequencies, and denoted there $M$) and the total mass ${\mathsf M}+m={\mathsf M}(1+q)$ (which
is used in PN/EOB works to adimensionalize frequencies, and denoted there $M$ as well). 
\begin{table}[htb] 
\begin{tabular}{l|c|c|c}
\hline\hline
\mbox{} & This paper & GSF literature (e.g., \cite{SBD,Barack:2010tm}) & PN/EOB literature (e.g., \cite{Damour:2009sm})\\
\hline\hline
particle mass & $m$ & $\mu$ & $m_1$ \\
black hole mass & ${\mathsf M}$ & $M$  & $m_2$ \\
total mass &${\mathsf M}+m$ & -- & $M\equiv m_1+m_2$ \\
``small'' mass ratio & $q\equiv m/{\mathsf M}$ & -- & -- \\
symmetric mass ratio & $\nu\equiv m {\mathsf M}/(m + {\mathsf M})^2$ & --& $\nu\equiv m_1m_2/(m_1+m_2)^2$ \\
effective mass & $\mu=m {\mathsf M}/(m + {\mathsf M}) $ & -- & $\mu\equiv m_1m_2/(m_1+m_2)$ \\
\hline\hline
\end{tabular}
\caption{Our notation for various mass quantities, compared with the common GSF notation and with the standard PN or EOB notation. Readers should be wary of these notation differences, which can bring confusion. 
}
\label{table:masses}
\end{table}

\subsection{GSF treatment}

\subsubsection{GSF-corrected circular orbits}

In GSF theory the binary dynamics is described perturbatively: At the limit $m\to 0$ the orbit is a geodesic of the ``background'' Schwarzschild geometry of mass ${\mathsf M}$, and we consider the $O(q)$ perturbation to this geodesic due to the effect of the GSF. Here we focus on the {\it conservative} effect alone, and ignore dissipation. We start by considering {\it circular} orbits, parameterized by their (Schwarzschild) radius $r=r_0={\rm const}$. (Here and in the following, $r_0$ denotes the radius of the perturbed, i.e., GSF-corrected circular orbit. Such
orbits exist because we are considering here only the conservative part of the GSF.)  Without loss of generality we let the orbit lie in the equatorial plane. The Schwarzschild components of the particle's (GSF-corrected) four-velocity $u^{\alpha}\equiv d x^{\alpha}/d\tau$
($\tau$ denoting the background proper time along the perturbed orbit) can be written as
\begin{equation}
u^{\alpha}=\frac{Er_0}{r_0-2{\mathsf M}}\left\{1,0,0,\Omega_{\varphi}\right\},
\end{equation}
where $E\equiv - g_{t \alpha}u^{\alpha}$ is the specific 
(background-defined) `energy' of the particle, and 
$\Omega_{\varphi}\equiv d \varphi/dt$ is the azimuthal coordinate-time ($\varphi$-)frequency.
The perturbed values of the latter two quantities (squared, for convenience) are related, through $O(q)$, to the radius $r_0$
of the orbit by \cite{BS}
\begin{equation} \label{E}
E^2=\frac{(r_0-2{\mathsf M})^2}{r_0(r_0-3{\mathsf M})}\left[1-\frac{r_0^2}{m(r_0-2{\mathsf M})} F_{\rm circ}^r\right],
\end{equation}
\begin{equation}\label{Omegaphi}
\Omega_{\varphi}^2=\frac{{\mathsf M}}{r_0^3}\left[1-\frac{r_0^2(r_0-3{\mathsf M})}{m{\mathsf M}(r_0-2{\mathsf M})}F_{\rm circ}^r\right],
\end{equation}
where $F^r_{\rm circ}(\propto m^2)$ is the radial component of the GSF acting on the particle (which, in the case of circular motion, is entirely conservative). We note that, in practice, it is sufficient for us to evaluate the GSF along the background geodesic rather than along the perturbed orbit, as the resulting difference in the value of dynamical quantities like $E$ or $\Omega_{\varphi}$ is only $O(q^2)$ and can be neglected here. 

We recall that in GSF theory the radius $r_0$ is gauge dependent, just like the GSF itself \cite{Barack:2001ph}. In this work we will always consider the GSF in the {\em Lorenz gauge}, which is the choice of gauge made in Refs.\ \cite{BS,Barack:2010tm}. The GSF component $F^r_{\rm circ}$, and all other GSF quantities introduced below, should be understood to be given specifically in the Lorenz gauge.  We also recall that in GSF theory the value of the particle's specific energy $E$, given in Eq.\ (\ref{E}), is dependent upon the gauge. However, the frequency $\Omega_{\varphi}$ in Eq.\ (\ref{Omegaphi}) is gauge invariant. More precisely, $\Omega_{\varphi}$ is invariant under the restricted class of $O(q)$ gauge transformations whose displacement vectors  respect the helical symmetry of the perturbed spacetime (see, e.g., \cite{SBD} for a detailed discussion of this point). 

Importantly, however---as discussed previously in the literature \cite{Barack:2005nr,SBD,Damour:2009sm}---the gauge transformation relating the Lorenz-gauge metric perturbation to the ones used in PN and EOB studies, does not fall in the above category, and therefore does not leave $\Omega_{\varphi}$ invariant. Nonetheless, it is straightforward to account for this gauge difference using what amounts to a simple $O(q)$ ``rescaling'' of the time $t$ \cite{SBD},
\begin{equation}
t\to \hat{t}=(1+q\alpha)t,
\end{equation}
with 
\begin{equation}\label{alpha}
\alpha={\mathsf M}[r_0(r_0-3{\mathsf M})]^{-1/2}.
\end{equation}
This leads to a ``rescaled-$t$'' frequency, given through $O(q)$ by
\begin{equation}\label{Omegaphihat}
\hat\Omega_{\varphi}= (1-q\alpha)\Omega_{\varphi}.
\end{equation}
The frequency $\hat\Omega_{\varphi}$ refers to an asymptotically-flat coordinate system (as the one employed in PN and EOB studies) and it therefore provides a useful reference point for comparison between GSF and PN/EOB calculations. To make further contact with standard PN/EOB notions, we also introduce the dimensionless gravitational potential $x$ (``total mass over radius''), defined through
\begin{equation}\label{x}
x\equiv[({\mathsf M}+m)\hat\Omega_{\varphi}]^{2/3}.
\end{equation}
The  quantity $x$ is the primary gauge-invariant characteristic of the conservative
dynamics of GSF-corrected circular orbits.
Using Eq.\ (\ref{x}) in conjunction with Eqs.\ (\ref{Omegaphi}), (\ref{alpha}) and (\ref{Omegaphihat}), one finds that the relation between $r_0$ and $x$ is given through $O(q)$ by
\begin{equation}\label{r0}
r_0=\frac{\mathsf M}{x}\left[1+\frac{2}{3}q\left(1-\frac{x}{(1-3x)^{1/2}}\right)-\frac{1-3x}{3 x^2(1-2x)q}\,F^r_{\rm circ}\right],
\end{equation}

\subsubsection{GSF-corrected slightly-eccentric orbits}

A second gauge-invariant quantity associated with the conservative dynamics can be constructed by considering a small-eccentricity perturbation of the circular orbit. Through linear order in the eccentricity $e$, the radius of the slightly-eccentric orbit can be written in the form $r(\tau)=r_0(1-e\cos\omega_{r} \tau)$, where $\omega_r$ is the radial frequency defined with respect to the proper-time $\tau$ along the orbit (and where, without loss of generality, we have set the orbital phase so that $\tau=0$ corresponds to a periastron passage). Henceforth, $r_0$
will denote the average radius of such a slightly-eccentric orbit, and we shall associate the gauge invariants $\hat\Omega_{\varphi}$ and $x$ introduced above to the `mean' circular orbit corresponding to the eccentric orbit in question.

As shown in Ref.\ \cite{ISCOLetter}, the various non-zero components of the conservative GSF along a slightly-eccentric orbit have the form [through $O(e)$]
\begin{eqnarray}
F^r&=&F^r_{\rm circ}+eF^r_{1}\cos\omega_{r} \tau, \\
F_{t}&=& e\omega_{r} F_{t1}\sin\omega_{r} \tau,   \\
F_{\varphi}&=& e\omega_{r} F_{\varphi 1}\sin\omega_{r} \tau. 
\end{eqnarray}
An expression for $\omega_{r}$ at the limit $e\to 0$ and through $O(q)$ was obtained in Ref.\ \cite{ISCOLetter}. It reads
\begin{equation}\label{omegar}
\omega^2_{r}=\frac{{\mathsf M}(r_0-6{\mathsf M})}{r_0^3(r_0-3{\mathsf M})}
-\frac{3(r_0-4{\mathsf M})}{mr_0(r_0-3{\mathsf M})}F_{\rm circ}^r +\frac{1}{mr_0} F_1^r-\frac{2[{\mathsf M}(r_0-3{\mathsf M})]^{1/2}}{mr_0^4} F_{\varphi 1}.
\end{equation}
The first term on the right-hand side of this expression would give the value of the radial frequency in the test-particle limit $q \to 0$; the subsequent terms [each of $O(q)$] describe conservative GSF corrections.
For our current analysis we also introduce the radial frequency $\Omega_r$ defined with respect to coordinate time $t$, along with its ``rescaled-$t$'' version $\hat\Omega_r$. The latter is related to $\omega_r$ (in the limit $e\to 0$) through
\begin{equation}\label{Omegardef}
\hat\Omega_r=\frac{d\tau}{dt}\frac{dt}{d\hat{t}}\,\omega_r=\left[\frac{1-2M/r_0}{E(1+q\alpha)}\right]\omega_r.
\end{equation}
Substituting from Eqs.\ (\ref{E}) and (\ref{omegar}) we obtain, at the limit $e\to 0$ and through $O(q)$,
\begin{equation}\label{Omegarhat}
\hat\Omega^2_{r}=\frac{{\mathsf M}(r_0-6{\mathsf M})}{r_0^4}(1-2q\alpha)
+\frac{r_0-3{\mathsf M}}{mr_0^2}\left[
-\frac{3r_0-10{\mathsf M}}{r_0-2{\mathsf M}}F_{\rm circ}^r + F_1^r-\frac{2[{\mathsf M}(r_0-3{\mathsf M})]^{1/2}}{r_0^3} F_{\varphi 1}
\right].
\end{equation}

Since both frequencies $\hat\Omega_\varphi$ and $\hat\Omega_r$ are gauge invariant (in the aforementioned sense), the functional relation $\hat\Omega_r(\hat\Omega_\varphi)$, or equivalently $\hat\Omega_r(x)$, is a genuinely gauge-invariant characteristic of the conservative dynamics. Following Ref.\ \cite{Damour:2009sm}, we introduce here the more convenient (dimensionless) quantity 
\begin{equation}
W\equiv \left(\hat\Omega_r/\hat\Omega_\varphi\right)^2,
\end{equation}
and consider the equivalent gauge-invariant relation $W(x)$. With Eqs.\ (\ref{Omegaphi}), (\ref{Omegaphihat}), (\ref{r0}) and (\ref{Omegarhat}), this relation takes the form 
\begin{equation}\label{Wexp}
W(x)=1-6x+ q\rho(x)+O(q^2),
\end{equation}
where the $O(q)$ part is given by
\begin{equation}\label{rhoSF}
\rho(x)= f_{r0}(x) \tilde F_{\rm circ}^r+f_{r1}(x) \tilde F_1^r+f_{\varphi 1}(x) \tilde F_{\varphi 1}+f_{(\alpha)}(x).
\end{equation}
Here $\tilde F_{\rm circ}^r\equiv q^{-2} F_{\rm circ}^r$, $\tilde F_1^r\equiv q^{-2} F_1^r$ and
$\tilde F_{\varphi 1}\equiv m^{-2}F_{\varphi 1}$, and the various $x$-dependent
coefficients read
\begin{eqnarray}\label{fr0}
f_{r0}(x)&=&-\frac{2(1-3x)(1-x)}{x^2(1-2x)},
\\
f_{r1}(x)&=&\frac{1-3x}{x^2},
\\
f_{\varphi 1}(x)&=&-2x^{1/2}(1-3x)^{3/2},
\\ \label{falpha}
f_{(\alpha)}(x)&=&4x-\frac{4x^2}{(1-3x)^{1/2}}.
\end{eqnarray}

Provided with GSF data for slightly-eccentric orbits, Eq.\ (\ref{rhoSF}) can be used to compute the GSF-induced shift in the invariant function $W(x)$. Let us note that this function has a simple physical interpretation: The quantity
\begin{equation}
k\equiv\hat\Omega_\varphi/\hat\Omega_r-1=W^{-1/2}-1
\end{equation}
describes the fractional periastron advance per radial period. The $O(q)$ GSF correction to $k$ is given by
\begin{equation}
\delta k=-\frac{1}{2}(1-6x)^{-3/2}q \rho(x).
\end{equation}
Note that, in the test-mass limit, we have the well known ISCO behavior $k\sim  (1-6x)^{-1/2}$. This singular behavior is avoided by working (as we do here) with the function $W(x)=(1+k)^{-2}$, which is smooth across the ISCO (and vanishes there); cf.\ \cite{Damour:1999cr,Damour:2000we}.

It should also be noted that the quantity $\rho(1/6)$ is related in a simple way to the ISCO frequency shift computed in Ref.\ \cite{ISCOLetter}. The ISCO location is defined through $W(x_{\rm isco})=0$, which, recalling Eq.\ (\ref{Wexp}), gives $q\rho(1/6)=6x_{\rm isco}-1$ [neglecting terms of $O(q^2)$]. Using Eqs.\ (\ref{x}) and (\ref{Omegaphihat}) this then leads to 
\begin{equation}
\rho(1/6)=\frac{2}{3}\left[q^{-1}\Delta\Omega_{\rm isco}/\Omega_{\rm isco}+ 1 -1/\sqrt{18}\right],
\end{equation}
where $\Delta\Omega_{\rm isco}/\Omega_{\rm isco}$ is the fractional $O(q)$ shift in the (Lorenz-gauge) $\Omega_{\varphi}$ at the ISCO. 
The numerical value obtained in \cite{ISCOLetter} for the latter was $(0.4870\pm0.0006)q$, giving
$\rho(1/6)=0.8342\pm 0.0004$. As part of our current analysis we will obtain the more accurate value  
$\rho(1/6)=0.83413\pm 0.00004$ (cf.\ Table \ref{Table:data} below).


\subsection{EOB treatment}

An EOB treatment of slightly-eccentric orbits in Schwarzschild in the small mass-ratio case was presented in the recent work \cite{Damour:2009sm}. Here we shall merely summarize the relevant results, and we refer to reader to Ref.\ \cite{Damour:2009sm} for full details. 

In the EOB approach the two components of the binary system are treated ``on equal footing'': Unlike in the GSF approach, where one speaks of a ``background'' and a ``perturbation'', here the conservative dynamics is described in terms of an effective geometry which depends on the binary masses only through the symmetric combinations $\mu\equiv {\mathsf M}m/({\mathsf M}+m)$ (effective mass) and $\nu\equiv {\mathsf M}m/({\mathsf M}+m)^2$ (symmetric mass ratio; cf.\ Table \ref{table:masses}). The effective metric $g_{\mu\nu}^{\rm eff}$ is spherically symmetric, 
and is a `deformed' version of the Schwarschild metric (of mass ${\mathsf M}+m$), with $\nu$ being the deformation parameter. When using Schwarschild-like EOB coordinates, $g_{\mu\nu}^{\rm eff}$ is fully described by two functions of the EOB radial coordinate $r_{\rm EOB}$,
namely $A(u;\nu) = - g_{00}^{\rm eff}$ and 
$\bar D(u;\nu)= -(g_{00}^{\rm eff}g_{rr}^{\rm eff})^{-1}$. Here, the argument
$u\equiv ({\mathsf M}+m)/r_{\rm EOB}$ is the dimensionless EOB gravitational potential, which conveniently parametrizes the binary separation in the EOB formalism. As already mentioned
above, the full EOB description of the conservative dynamics involves, besides $A(u;\nu)$
and $\bar D(u;\nu)$, a third function, which depends not only on $u$ and $\nu$ but also
on the radial (relative) momentum $p_r$, namely $Q(u,p_r;\nu)$. However, as explained in
Ref.\ \cite{Damour:2009sm}, the conservative EOB dynamics of {\em small-eccentricity} orbits
depends only on the two functions $A(u;\nu)$ and $\bar D(u;\nu)$ parametrizing the 
effective metric $g_{\mu\nu}^{\rm eff}$.

To make contact with GSF theory, one formally expands the functions $A(u;\nu)$ and $\bar D(u;\nu)$ in powers of the symmetric mass ratio through $O(\nu)$ (noting $\nu=q$ through this order). This gives \cite{Damour:2009sm}
\begin{eqnarray}
A(u,\nu)&=&1-2u+\nu a(u) +O(\nu^2)\nonumber\\
        &=&1-2u+q a(u) +O(q^2) , \\
\bar D(u,\nu)&=&1+\nu \bar d(u) +O(\nu^2)\nonumber\\
			 &=&1+q \bar d(u) +O(q^2),
\end{eqnarray} 
where the functions $a(u)$ and $\bar d(u)$ carry, in principle, all information about the conservative GSF along slightly eccentric orbits. [The GSF dynamics of large-eccentricity
orbits would also involve a third function, say $q(u,p_r)$.] In particular, it  
was shown in Ref.\ \cite{Damour:2009sm} that the quantity $\rho(x)$ defined in Eq.\ (\ref{Wexp}) can be constructed from these two EOB functions using the formula 
\begin{equation}\label{rhoEOB}
\rho(x)=\rho_{E}(x)+\rho_a(x)+\rho_d(x),
\end{equation}
where 
\begin{eqnarray}
\rho_{E}(x)&=&4x\left(1-\frac{1-2x}{\sqrt{1-3x}}\right), \\
\rho_{a}(x)&=&a(x)+x a'(x)+\frac{1}{2}x(1-2x)a''(x), \\
\rho_{d}(x)&=&(1-6x)\bar d(x), \label{rhoEOBpieces}
\end{eqnarray}
with a prime denoting $d/dx$. [Although we shall not need it here, we note that the value
of the EOB gravitational potential $u$ along circular orbits is related to the gauge-invariant frequency 
parameter $x$ introduced above via a relation of the type $u=x+O(\nu)$; the $O(\nu)$ difference between $u$ and $x$ is given explicitly in Eq.\ (4.22) of \cite{Damour:2009sm}.] 
Later in our discussion we shall occasionally refer to the `corrected' $\rho$ function obtained by subtracting from it the contribution $\rho_{E}(x)$. We shall then denote it as
\begin{equation}\label{rhotilde}
\tilde \rho(x)\equiv \rho(x) -\rho_{E}(x)= \rho_a(x)+\rho_d(x).
\end{equation}

In summary, a GSF computation of the function $\rho(x)$ gives a direct access to a linear combination of the $O(\nu)$ EOB functions $a(x)$ and $\bar d(x)$, and their derivatives. 

\subsubsection{PN expansion}

While, as we just saw, GSF theory can give a handle on the {\em strong-field} \footnote{In principle, GSF methods could explore the function $\rho(x)$ in the full interval $0<x<1/3$ where
circular geodesic orbits exist. However, the present structure of our GSF code makes it impossible
to explore the perturbations of the unstable circular orbits below the ISCO, i.e., in the range
$1/6<x<1/3$.} behavior of the EOB functions, PN theory gives access  to the {\em weak-field}
behavior of the functions $a(u)$ and $\bar d(u)$, i.e., to their expansions in powers of $u$.
 The structure of the PN expansions of these two functions reads
\begin{equation} \label{adexp}
a^{\rm PN}(u)=\sum_{n\geq 3} a_n u^n, \quad\quad \bar d^{\rm PN}(u)=\sum_{n\geq 2} \bar d_n u^n,
\end{equation}
where the (2PN and 3PN) coefficients $a_3$ and $a_4$, as well as $\bar d_2$ and $\bar d_3$, are pure numbers, while higher-order coefficients generally involve a logarithmic dependence on $u$. Inserting the
expansions (\ref{adexp}) (with the replacement $u \to x$ \footnote{This replacement is only a change in the name of a mathematical argument. It differs from the physical
statement that $u=x+O(\nu)$ along the sequence of circular orbits.}) into Eqs.\ (\ref{rhoEOB})--(\ref{rhoEOBpieces}) yields the weak-field ($x \to 0$), or PN, expansion of the function $\rho(x)$, say
\begin{equation}
\rho^{\rm PN}(x)=\sum_{n\geq 2} \rho_n x^n.
\end{equation}
Note that a term of order $O(x^n)$ in this expansion corresponds to the $n$th PN order [while
the $n$PN level corresponded to a term of order $O(u^{n+1})$ in the expansion of the function $a(u)$].
The coefficients of the first two terms in the expansion $\rho^{\rm PN}(x)$, corresponding to
the 2PN and 3PN levels, are pure numbers, while the higher-order coefficients $\rho_n$, for $n=4,5,\ldots$ inherit a logarithmic dependence on $x$ from that of 
$a_5,a_6,\ldots$ and $\bar d_4,\bar d_5,\ldots$.
We shall then write the PN expansion of $\rho(x)$ in the explicit form 
\begin{equation}\label{rhoPN}
\rho^{\rm PN}(x)=\rho_2 x^2+\rho_3 x^3 + (\rho_4^{\rm c}+\rho_4^{\rm log}\ln x)x^4+(\rho_5^{\rm c}+\rho_5^{\rm log}\ln x)x^5+O(x^{6+0}),
\end{equation}
where $\rho_4^{\rm c}$, $\rho_4^{\rm log}$, $\rho_5^{\rm c}$ and $\rho_5^{\rm log}$ are all numerical coefficients, and where the symbol $O(x^{6+0})$ refers to the presence of 
logarithmic corrections (of a finite, but unspecified, order) in the $O(x^6)$ remainder. [As argued in \cite{Damour:2009sm,Blanchet:2010zd}  one expects no higher powers of $\ln x$ to occur through $O(x^5)$, although such terms may well appear at higher orders.]

The first two terms in the PN expansion (\ref{rhoPN})
are determined by the currently known
3PN results. More precisely, inserting the 3PN results \cite{Damour:2000we}
\begin{eqnarray}
a_3&=&2, \label{a3} \\ 
a_4&=&\frac{94}{3}-\frac{41\pi^2}{32}\simeq 18.687903, \label{a4}\\
\bar d_2&=&6, \label{d2}\\ 
\bar d_3&=&52 \label{d3}
\end{eqnarray}
into the first two equations obtained by inserting the
expansions (\ref{adexp})  into Eqs.\ (\ref{rhoEOB})--(\ref{rhoEOBpieces}), namely
\begin{eqnarray}
\rho_2 &=& 2+3a_3+\bar d_2, \\
\rho_3 &=& -\frac{3}{2}-2a_3+6a_4-6\bar d_2+\bar d_3, 
\end{eqnarray}
yields
\begin{eqnarray}\label{rho23}
\rho_2&=&14, \nonumber\\
\rho_3&=& \frac{397}{2}-\frac{123}{16}\pi^2 \simeq 122.627416.
\end{eqnarray}
Recently, the logarithmic contributions to $\rho(x)$  at
the 4PN and 5PN levels, i.e., the analytic values of $\rho_4^{\rm log}$ and $\rho_5^{\rm log}$,
have been derived by one of us \cite{DamourLogs}, using effective-action techniques, with the results
\begin{eqnarray}\label{rho45log}
\rho_4^{\rm log} &=& \frac{16}{15}\times 157\simeq 167.466666, \nonumber\\
\rho_5^{\rm log} &=&-\frac{11336}{7}\simeq -1619.428571. 
\end{eqnarray}
By contrast, the values of the non-logarithmic coefficients $\rho_4^{\rm c}$ and $\rho_5^{\rm c}$ remain unknown, being beyond the present capabilities of PN theory.

A GSF computation of $\rho(x)$ can be tested against the PN expression (\ref{rhoPN}) (with the known values of $\rho_2$, $\rho_3$, $\rho_4^{\rm log}$ and $\rho_5^{\rm log}$), and, furthermore, can in principle be used to determine $\rho_4^{\rm c}$, $\rho_5^{\rm c}$, and possibly some other, higher-order terms in the expansion. For this purpose, it is useful to examine the PN behavior of the GSF expression for $\rho(x)$, Eq.\ (\ref{rhoSF}). Expanding the coefficients in Eqs.\ (\ref{fr0})--(\ref{falpha}) in powers of $x$, we have
\begin{eqnarray}\label{rhoSFPN}
\rho(x)&=& \left(-2x^{-2}+4x^{-1}+2+4x+8x^2+\ldots\right) \tilde F_{\rm circ}^r 
\nonumber\\
&& +         \left(x^{-2}-3x^{-1}\right) \tilde F_1^r 
\nonumber\\
&& +
x^{1/2}\left(-2+9x-27x^2/4+\ldots\right) \tilde F_{\varphi 1} 
\nonumber\\
&&
+ 4x-4x^2-6x^3+\ldots.
\end{eqnarray}
Note that, when $ x \to 0$, Eq.\ (\ref{rhoPN}) says that $\rho(x)$ vanishes proportionally to
$x^2$, while the first two terms in the GSF expression Eq.\ (\ref{rhoSFPN}) involve 
coefficients that blow up proportionally to $x^{-2}$. This means that there must exist
delicate cancellations between the various terms on the right-hand-side of Eq.\ (\ref{rhoSFPN}).
It is well-known (see, e.g., \cite{Detweiler:2003ci}) that the leading weak-field contribution to the GSF is of {\em Newtonian} origin, and simply comes from the `recoil' of the large mass with respect to
the center of mass of the binary system. This leads to a leading-order GSF given by
${\bf F}^{\rm (leading)}= + 2 m^2 {\bf r}/r^3$, i.e., a purely radial GSF, which [upon inserting
$r(\tau)=r_0(1-e\cos\omega_{r} \tau)$] yields $\tilde F_{\rm circ}^{r{\rm (leading)}}= 2x^2$
and $\tilde F_{1}^{r{\rm (leading)}}= 2 \tilde F_{\rm circ}^{r{\rm (leading)}} = 4 x^2$. Inserting
these results in the above expression for $\rho(x)$, one finds that the first line on 
the right-hand side of Eq.\ (\ref{rhoSFPN}) contributes $-4 + O(x)$, while the
second line contributes $ + 4 + O(x)$. As expected we have a cancellation of the leading-order 
terms, but this cancellation could leave a contribution to $\rho(x)$ of order $O(x)$, i.e.,
of 1PN. To show analytically that the 1PN terms also cancel out
in $\rho(x)$ would require a 1PN-accurate analytic expression for the GSF in the
Lorenz gauge. Such analytical knowledge is unfortunately not yet available, especially
for generic, eccentric orbits.
(For {\em circular} orbits, the first few terms in the PN expansion of the
Lorenz-gauge GSF have been estimated from a fit to numerical data---see Eq.\ (56) in \cite{BS}.)
That the above cancellation at 1PN does actually occur will be verified numerically below. 

\section{Numerical method and results}

In this section we describe the numerical computation of $\rho(x)$ using GSF methods, and present the raw data coming out of this calculation. In subsequent sections we will analyze these data and explore what can be learned from a comparison with EOB predictions.

We recall, noting Eq.\ (\ref{rhoSF}), that a computation of $\rho(x)$ requires three pieces of input, namely the GSF coefficients $\tilde F^r_{\rm circ}$, $\tilde F^r_1$ and $\tilde F_{\varphi 1}$. Recall also that $\tilde F^r_{\rm circ}$ is related to the radial (conservative) component of the GSF for a strictly circular orbit, while $\tilde F^r_1$ and $\tilde F_{\varphi 1}$ are related to the small-eccentricity perturbations in the radial and azimuthal components of the conservative GSF. Section V.B of Ref.\ \cite{Barack:2010tm} described two independent strategies for determining these coefficients. In the first strategy (dubbed {\it method I}) one considers a sequence of geodesic orbits with decreasing eccentricities, which approach the desired circular orbit along a suitable ``track'' in the 2-dimensional parameter space of eccentric geodesics. One then computes the GSF along each orbit in the sequence, and the necessary GSF coefficients are obtained through extrapolation to zero eccentricity (see \cite{Barack:2010tm} for details). In the second strategy ({\it method II}) the time-domain field equations themselves are expanded in the eccentricity parameter $e$ through $O(e)$, with the coefficient $\tilde F^r_{\rm circ}$ then obtained from the $O(e^0)$ set of equations and the coefficients $\tilde F^r_1$ and $\tilde F_{\varphi 1}$ obtained from the $O(e^1)$ ones. While {\it method II} is somewhat more difficult to implement, it is also significantly more computationally efficient, and allows one to obtain the necessary GSF coefficients with a greater accuracy. In Ref.\ \cite{Barack:2010tm} (as part of the ISCO shift analysis) both methods were implemented in order to obtain the three GSF coefficients at the single radius $r_0=6{\mathsf M}$ ($x=1/6$). In that computation, {\it method II} was incorporated to obtain a high-accuracy result, with {\it method I} used to confirm that result. 

Here we shall use {\it method II} in order to obtain the GSF coefficients---and thereby $\rho$---for a large sample of $x$ values between the ISCO ($x=1/6$) and the weak-field radius $r_0=80{\mathsf M}$ ($x=0.0125$). The precise implementation procedure follows closely that of Ref.\ \cite{Barack:2010tm} and is based on the time-domain Lorenz-gauge GSF code described therein. We list below several minor details in which our current analysis deviates from that of Ref.\ \cite{Barack:2010tm}.

First, we note that the computation of the GSF coefficients at the ISCO in Ref.\ \cite{Barack:2010tm} involved a certain extrapolation procedure even within {\it method II}. This is because the source terms in the $O(e)$ field equations become divergent as the radial frequency of the orbit approached zero [cf.\ Eqs.\ (F3)--(F12) in Appendix F of \cite{Barack:2010tm}], which makes it impractical to solve the $e$-perturbed equations at the ISCO itself. This additional computational burden is spared from us here: for each value $x<1/6$ we need only solve the set of $e$-perturbed field equations once, for a particle at precisely the desired orbit. However, we also reproduce here (with better accuracy) the ISCO values of the GSF coefficients, and for this we use an extrapolation procedure similar to that employed in Ref.\ \cite{Barack:2010tm}.

On the other hand, our task here is made more challenging by the need to consider relatively large orbital radii (these are particularly interesting for the purpose of comparing with PN results), because the computational cost of our time-domain evolution tends to grow fast with increasing radius. To understand the reason, we note that in time-domain computations such as that of Ref.\ \cite{Barack:2010tm} one does not (and usually {\em can} not) impose accurate initial conditions for the numerical evolution; instead, one simply let any spurious waves resulting from the imperfection of the initial data ``dissipate away'' over time, making sure that the time-evolution proceeds long enough for the magnitude of these transient waves to fall below a set threshold. The initial, transient stage of the evolution is then discarded, and one records only the physical, stationary late-time solution. The ``relaxation'' time of the spurious transient thus dictates the necessary evolution time (and we note that in our particular 1+1-dimensional implementation, the actual computational time is {\em quadratic} in the evolution time). Unfortunately, initial spurious waves from larger radii take longer to die off \footnote{This is a combined result of (1) initial radiation from the particle having to travel longer before being scattered off the strong-field potential barrier surrounding the black hole, and (2) the fact that the amplitude of residual late-time decay tails increases with decreasing frequency, as low-frequency radiation is less susceptible to black-hole absorption.}, and hence necessitate a longer evolution. At large orbital radii ($x\lesssim 1/50$) this becomes the main limiting factor in our numerics.

In our numerical implementation we have adjusted the evolution time as a function of $x$ in order to ensure that the error from residual initial waves is kept below (or at most comparable to) other sources of numerical error (we refer the reader to Ref.\ \cite{Barack:2010tm} for a detailed discussion of the various sources of error in our computation and of how these are monitored and controlled). Fortunately, it is the case that higher multipole modes of the metric perturbation decay faster than lower ones, which is easily exploitable in a mode-sum treatment like ours: It allowed us to reduce the evolution time for higher multipole numbers without compromising the numerical accuracy. Through experimentation, we arrived at the following practical scheme for determining the evolution time: (i) For  $1/20\leq x< 1/6$ evolve the multipoles $l=0$--$3$ for $t=1000{\mathsf M}$  and higher multipoles for $t=500{\mathsf M}$. (ii) For $1/50\leq x <1/20$ evolve the multipoles $l=0$--$5$ for $t=1500{\mathsf M}$ and higher multipoles for $t=500{\mathsf M}$. (iii) For $x<1/50$ evolve the multipoles $l=0$--$5$  for 2000M and higher multipoles for $t=500{\mathsf M}$.

\begin{figure}[htb]
\includegraphics[width=12cm]{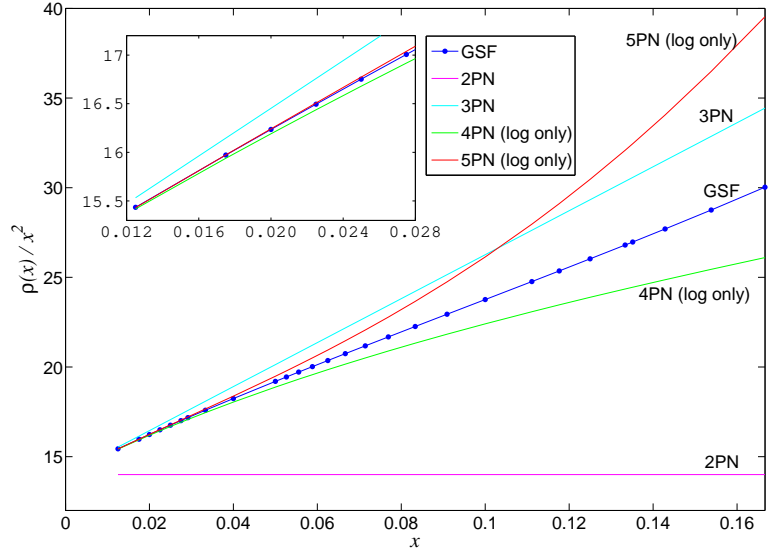}
\caption{Numerical GSF data for $\rho(x)$ compared with various EOB/PN approximations. For clarity we show $\rho(x)/x^2$ rather than $\rho(x)$ itself, recalling the small-$x$ asymptotic behavior $\rho(x)\propto x^2$. Dots represent the GSF data points shown in Table \ref{Table:data}, and the blue (darkest) line is a simple cubic spline interpolation of these data. Numerical error bars are too small to show on this scale. Other lines display various analytic PN models $\rho^{\rm PN}$ constructed from Eq.\ (\ref{rhoPN}); the plots labelled `nPN' show $\rho^{\rm PN}$ through $O(x^n)$, where at 4PN and 5PN only the known, logarithmic terms are included. The inset shows an expansion of the small-$x$ portion of the plot. Recall $x$ is the (dimensionless) gravitational potential, so $x\to 0$ corresponds to $r_0\to\infty$. The $x$-domain here extends to the ISCO location at $x=1/6$. 
}
\label{fig:rho}
\end{figure}
\begin{figure}[htb]
\includegraphics[width=10cm]{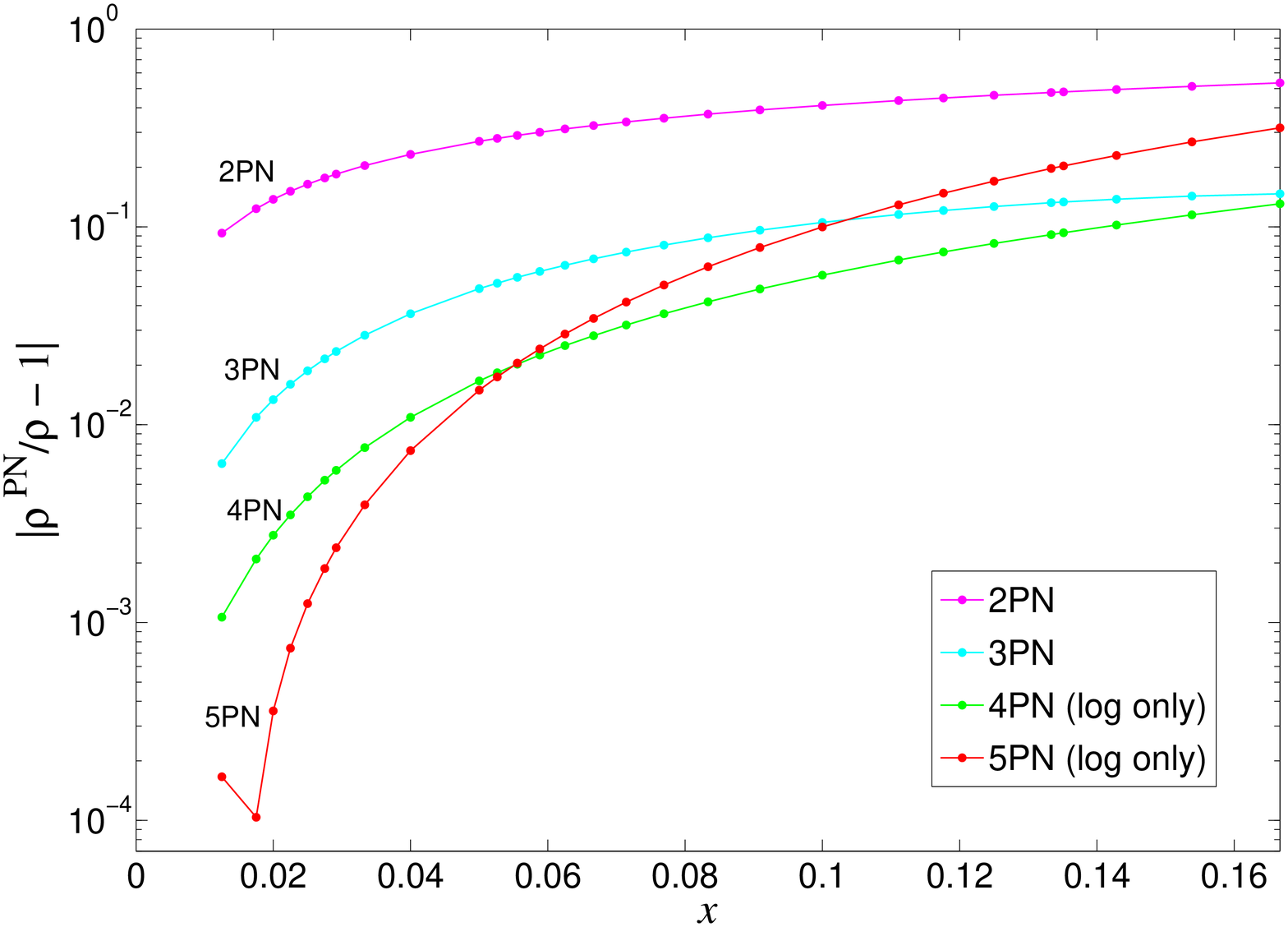}
\caption{The fractional difference $\left|\rho^{\rm PN}-\rho\right|/\rho$ between the various PN models shown in Figure \ref{fig:rho} and the numerical GSF data, as a function of $x$. (As in Figure \ref{fig:rho}, the 4PN and 5PN models include the logarithmic terms only.) Note the logarithmic scale of the vertical axis. The magnitude of the ``kink'' at the far left end of the 5PN plot is well within the numerical error for the left-most data point (at $x=1/80$).
}
\label{fig:rhodiff}
\end{figure}

Table \ref{Table:data} summarizes our numerical results for $\rho(x)$, and in Figure \ref{fig:rho} we plot these data (dividing $\rho$ by $x^2$ for clarity). Figure \ref{fig:rho} also displays, for comparison, the various PN approximations $\rho^{\rm PN}(x)$ derived from Eq.\ (\ref{rhoPN}). The agreement between the numerical data and the PN results is made more evident in Figure \ref{fig:rhodiff}, where we plot the relative differences $\left|\rho^{\rm PN}-\rho\right|/\rho$ as  functions of $x$.
\begin{table}[Htb]
\begin{tabular}{l|l|l}
\hline\hline
$r_0/{\mathsf M}$  & $x={\mathsf M}/r_0$ & \ $\rho$  \\
\hline\hline
80 			 &   0.0125 		&  0.0024117(9)\\
57.142\ldots &   0.0175     	&  0.0048913(6) \\
50 			 &   0.0200 		&  0.006494(2) \\
44.444\ldots &   0.0225     	&  0.008351(2) \\
40			 &   0.0250 		&  0.010470(1)\\
36.363\ldots &   0.0275 		&  0.0128610(8) \\
34.2857		 &   0.0291\ldots   &  0.0146099(8) \\
30			 &   0.0333\ldots 	&  0.0195438(4) \\
25			 &   0.0400 		&  0.0291863(3)\\
20			 &   0.0500 		&  0.0479916(5) \\
19			 &   0.0526\ldots 	&  0.053862(3) \\
18			 &   0.0555\ldots 	&  0.060857(2) \\
17			 &   0.0588\ldots 	&  0.069279(4) \\
16			 &   0.0625 		&  0.079537(2) \\
15			 &   0.0666\ldots 	&  0.092199(3) \\
14			 &   0.0714\ldots 	&  0.108061(2) \\
13			 &   0.0769\ldots 	&  0.128280(3) \\
12			 &   0.0833\ldots 	&  0.154578(3) \\
11			 &   0.0909\ldots 	&  0.189605(3) \\
10			 &   0.100		 	&  0.237610(4) \\
9	 		 &   0.111\ldots 	&  0.305750(5) \\
8.5			 &   0.117\ldots 	&  0.351000(6) \\
8	 		 &   0.125 			&  0.406767(6) \\ 
7.5		     &   0.133\ldots 	&  0.47651(1) \\
7.4			 &   0.135\ldots    &  0.492527(7) \\
7			 &   0.142\ldots 	&  0.56528(1) \\
6.8			 &   0.147\ldots    &  0.607693(9) \\
6.5	 		 &   0.153\ldots 	&  0.68059(1) \\
6   		 &   0.166\ldots    &  0.83413(4) \\
\hline\hline
\end{tabular}
\caption{Numerical GSF data for $\rho(x)$. Figures in brackets are rough estimates of the absolute error in the last displayed digit (so, for example, the value in the first line stands for $0.0024117\pm 0.0000009$). These estimates include the error from the finite numerical mesh size (including error from the large $l$ tail approximation \cite{Barack:2010tm}), added in quadrature to an estimate of the error from residual spurious initial waves. The value for $r_0=6{\mathsf M}$ is obtained through an extrapolation based on 18 data points (not shown here) between $r_0=6.0005{\mathsf M}$ and $r_0=6.05{\mathsf M}$.
}
\label{Table:data}
\end{table}

Inspecting Figures \ref{fig:rho} and \ref{fig:rhodiff}, we can make the following preliminary observations.
(i) The numerical GSF data seems to be perfectly consistent with the analytic PN expressions.
(ii) At sufficiently small $x$ (large $r_0$) the PN series appears to converge to the ``exact'' GSF result. 
(iii) This convergence does not appear to be ``monotonic'', in the sense that the partial PN sums alternate between ``overshooting'' and ``undershooting'' the GSF result. [This can be contrasted with the case of $u^t(x)$, where a monotonic convergence is observed at least through 7PN \cite{Blanchet:2009sd,Blanchet:2010zd}.]
(iv) Close to the ISCO the PN series does not show a good convergence; going to higher-order PN does not necessary improve the accuracy of the PN approximation there.
(v) However, as expected, the PN series does extremely well at small $x$; for $x\lesssim 1/50$ the 5PN model approximates $\rho$ to within a few parts in $10^4$, even when the (yet unknown) 4PN and 5PN constant terms are neglected. 

In the next section we will take a closer look at the GSF data, and explore more quantitatively what can be learned from comparing it with the EOB predictions.

\section{Comparison of GSF results with EOB/PN predictions}\label{comparison}

\subsection{Preliminary tests}

We begin by checking that the numerical GSF results are consistent with the EOB/PN prediction (\ref{rhoPN}), with the specific analytic values of the coefficients given in Eqs.\ (\ref{rho23})
 and (\ref{rho45log}). To this end, let us consider the 4PN residual quantities 
\begin{eqnarray}\label{eq:Delta4}
\Delta_4 (x)&\equiv& \left(\rho(x)-\rho_2 x^2-\rho_3 x^3 \right)/x^4,
\nonumber\\
\Delta_4^+ (x) &\equiv& \left(\rho(x)-\rho_2 x^2-\rho_3 x^3 - \rho_4^{\rm log}x^4\ln x \right)/x^4,
\nonumber\\
\Delta_4^{++} (x)&\equiv& \left(\rho(x)-\rho_2 x^2-\rho_3 x^3 - \rho_4^{\rm log}x^4\ln x - \rho_5^{\rm log}x^5\ln x \right)/x^4,
\end{eqnarray}
where the coefficients $\rho_2$, $\rho_3$, $\rho_4^{\rm log}$ and $\rho_5^{\rm log}$ are those given in Eqs.\ (\ref{rho23}) and (\ref{rho45log}) above. 
Using the notation introduced in Eq.\ (\ref{rhoPN}) above, we can write the
small-$x$ expansion of these quantities as
\begin{eqnarray}\label{eq:Delta4asy}
\Delta_4(x) &=&\rho_4^{\rm c}+\rho_4^{\rm log}\ln x+(\rho_5^{\rm c}+\rho_5^{\rm log}\ln x)x+O(x^{2+0}), \\
\Delta_4^+(x) &=&\rho_4^{\rm c}+(\rho_5^{\rm c}+\rho_5^{\rm log}\ln x)x+O(x^{2+0}), \label{eq:Delta4pasy} \\
\Delta_4^{++}(x) &=&\rho_4^{\rm c}+\rho_5^{\rm c} x+O(x^{2+0}). \label{eq:Delta4pppasy}
\end{eqnarray}
Note that the residue $\Delta_4$ diverges logarithmically at $x\to 0$, while the ``improved'' residue $\Delta_4^+$ is finite at this limit but has a divergent derivative there. The ``further improved'' residue $\Delta_4^{++}$, however, is both continuous and differentiable at $x = 0^+$, with $\Delta_4^{++}(0)=\rho_4^{\rm c}$ and $(\Delta_4^{++})'(0)=\rho_5^{\rm c}$. In particular, the task of extracting the unknown PN coefficients $\rho_4^{\rm c}$ and $\rho_5^{\rm c}$ from the numerical data amounts to resolving the values of $\Delta_4^{++}(x)$ and its derivative at $x=0$. 

\begin{figure}[htb]
\includegraphics[width=14cm]{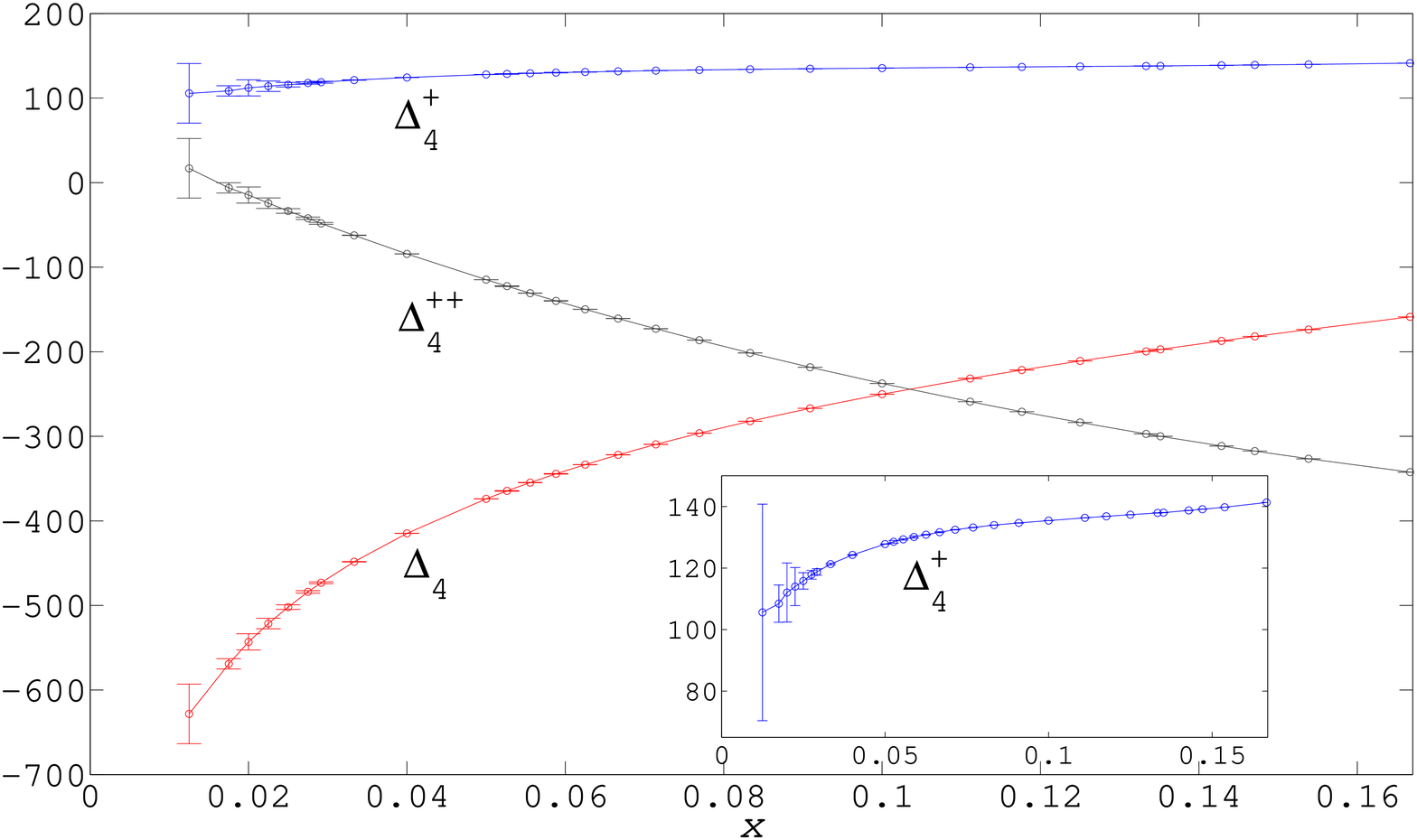}
\caption{Numerical GSF data for the 4PN residual quantities $\Delta_4(x)$, $\Delta_4^{+}(x)$ and $\Delta_4^{++}(x)$ [see Eqs.\ (\ref{eq:Delta4}) for definitions]. Curves are simple cubic-spline interpolations of the numerical data, and the inset displays the $\Delta_4^{+}$ data at a modified aspect ratio for better clarity. Error bars are computed from the estimated numerical errors in $\rho(x)$ indicated in Table \ref{Table:data}. As discussed in the text, these graphs illustrate the consistency of the numerical GSF data with the EOB/PN prediction through 3PN order. Furthermore, they suggest that the GSF shows the correct (analytically predicted) logarithmic running at 4PN and 5PN. 
}
\label{fig:Delta4}
\end{figure}

In Figure \ref{fig:Delta4} we have plotted the functions $\Delta_4(x)$, $\Delta_4^{+}(x)$ and $\Delta_4^{++}(x)$ based on the GSF numerical data. A visual inspection of the plot reveals the following. (i) The data for $\Delta_4(x)$ is consistent with the expected logarithmic divergence at $x\to 0$. (ii) This divergence seems to be eliminated in $\Delta_4^+(x)$, again as expected. A closer examination (see the inset) suggests that the derivative $(\Delta_4^+)'$ begins to vary rapidly at $x\lesssim 0.05$; although we cannot be conclusive, this behavior is what one would expect if $\Delta_4^+$ exhibited a $\propto x\ln x$ term as in Eq.\ (\ref{eq:Delta4pasy}). (iii) The data for the residue $\Delta_4^{++}$ seems to vary more smoothly as a function of $x$, even at small $x$; extrapolating ``by eye'' to $x\to 0$  suggests that both $\Delta_4^{++}$ and its derivative attain finite values at $x=0$. 

These results give a firmer basis to the conclusion 
that the GSF data is indeed consistent with the PN expression (\ref{rhoPN}), with
the specific analytical values of the 2PN and 3PN coefficients given in Eqs.\ (\ref{rho23}).  
In particular, our results confirm the occurrence of a high level of cancellation in the
small-$x$ expansion of the GSF expression for $\rho(x)$, Eq.\ (\ref{rhoSFPN}), resulting in
a $\propto x^2$ leading-order behavior in agreement with the prediction of PN theory \footnote{The 2PN nature of $\rho(x)$ is directly linked to the fact that the effective EOB metric of a binary
system starts to differ from the Schwarzschild metric only at 2PN \cite{Buonanno:1998gg}.}.
Furthermore, Figure \ref{fig:Delta4} suggests that the GSF data correctly capture the 
logarithmic terms occurring at 4PN and 5PN, which were recently determined analytically.

Figure \ref{fig:Delta4} also visually illustrates the difficulties inherent in our comparison. Even though the fractional numerical error in $\rho(x)$ has been kept roughly uniform ($x$-independent) in our computation, the corresponding error bars on the various $\Delta_4$ residues increase significantly with decreasing $x$ (as expected). This, of course, reduces our ability to make precise statements about the behavior of the GSF data at small $x$, particularly at sub-leading PN orders. The problem is exacerbated by the aforementioned numerical cancellation of the $O(x^0)$ and $O(x)$ terms in Eq.\ (\ref{rhoSFPN}), which effectively acts to amplify the numerical error in $\rho(x)$ at small $x$. To offset this problem it would be desirable, in principle, to adjust the accuracy goal set for $\rho(x)$ as a function of $x$, so that higher accuracy is sought for smaller $x$. However, this cannot be achieved easily with the current GSF code architecture: Recall that our computational cost increases quickly with decreasing $x$ as a result of the longer dissipation time of the initial spurious radiation. Likewise, it would be desirable to obtain accurate GSF data for $x$ values smaller than $1/80$, which would give us a better handle on the asymptotic PN behavior. This too, however, is extremely difficult to achieve with our current code for the same reason as above.

\subsection{Quantitative test of the GSF data against known EOB/PN terms}

Next, let us inspect the agreement between the GSF results and the analytic predictions of EOB/PN in a more quantitative fashion. We will do so by fitting PN models to the numerical $\rho(x)$ data and assessing the goodness of the fits using a standard $\chi^2$ analysis. To avoid the
loss of accuracy inherent in multi-dimensional fits \footnote{If we wanted to check 
{\em simultaneously} the numerical values of the 4 analytically computed PN parameters, 
$\rho_2$, $\rho_3$, $\rho_4^{\rm log}$, and $\rho_5^{\rm log}$, we would need to fit for 6 PN parameters at once, including the two extra non-logarithmic terms at 4PN and 5PN.}, we will be using a ``marginalization'' procedure whereby we fit only one PN parameter at a time, then fix the value of that parameter (in accordance with the analytic PN prediction) and proceed to inspect the next PN level. At each stage of this procedure we will fit our data against a few different PN models. For easy reference, we introduce the notation 
\begin{eqnarray}\label{PNmodels}
\rho^{\rm nPN}&:&\quad  \text{$\rho^{\rm PN}$ up to $O(x^n)$ inclusive, {\em excluding} the $n$PN log term}, \nonumber\\
\rho^{\rm nPN+}&:&\quad  \text{$\rho^{\rm PN}$ up to $O(x^{n+0})$ inclusive, i.e., {\em including} the $n$PN log term},
\end{eqnarray}
where ``$n$PN log term'' refers to a term of the form $\rho_n^{\rm log}x^n \ln x$. Our convention is that both $\rho^{\rm nPN}$ and $\rho^{\rm nPN+}$ include all logarithmic terms occurring at orders 4PN through to ($n-1$)PN. 

We begin with the leading 2PN term, $\rho\sim \rho_2 x^2$, to check to what extent our GSF
numerical data confirm the analytically predicted value of the 2PN coefficient:
$\rho_2^{\rm analytic}=14$. We first consider a set of PN models $\rho^{\rm 2PN},\ldots,\rho^{\rm 7PN}$ and $\rho^{\rm 4PN+},\ldots,\rho^{\rm 7PN+}$ with all PN coefficients taken as free fit parameters. We least-squares-fit each model to our numerical GSF data, weighed by the estimated numerical error from Table \ref{Table:data} (we do that in practice by minimizing the corresponding $\chi^2$ function over all fit parameters, using the built-in $\mathtt{Minimize}[\, ]$ function available in {\it Mathematica}). For each fit model we record the best-fit value of $\rho_2$, ignoring all other fit parameters for now. We also record the value of (the minimized) $\chi^2$, the value of the $L^{\infty}$-norm (i.e., the maximum magnitude of the absolute difference $\vert \Delta \rho \vert $ between a data point and the best-fit model), and the number of degrees of freedom for the fit (DoF; the number of numerical data points less the number of fit parameters). The results are presented in the upper part of Table \ref{Table:rho2}. At the bottom (last 3 lines) of Table \ref{Table:rho2} we present similar fit results for model $\rho^{\rm 7PN}$, where this time we fix some of the higher-order parameters $\rho_3$, $\rho_4^{\rm log}$, and $\rho_5^{\rm log}$ at their analytically-known values.
\begin{table}[Htb]
\begin{tabular}{lllll}
\hline\hline
fit model$\quad$ &  fixed params.~$\quad$ & $\rho_2$ (best fit)$\quad$ &  $\chi^2$/DoF $\quad$ & $L^{\infty}$-norm\\
\hline\hline
$\rho^{\rm 2PN}$ &	none &   $21.5941$ 	&  $6.8\times 10^7$ &  $2.3\times 10^{-1}$ \\
$\rho^{\rm 3PN}$ &	none &   $14.5748$ 	&  $5810$  			&  $3.6\times 10^{-3}$ \\
$\rho^{\rm 4PN}$ &	none &   $14.5135$ 	&  $5264$  			&  $4.8\times 10^{-3}$ \\
$\rho^{\rm 4PN+}$& 	none &   $13.9665$ 	&  $29.4$  		    &  $6.0\times 10^{-4}$ \\
$\rho^{\rm 5PN}$ &	none &   $14.0544$ 	&  $4.08$  		    &  $2.0\times 10^{-4}$ \\
$\rho^{\rm 5PN+}$& 	none &   $13.9721$ 	&  $0.74$  		    &  $4.5\times 10^{-5}$ \\
$\rho^{\rm 6PN}$ &	none &   $14.0106$ 	&  $0.59$  		    &  $1.6\times 10^{-5}$ \\
$\rho^{\rm 6PN+}$& 	none &   $13.9619$ 	&  $0.58$  		    &  $1.7\times 10^{-5}$ \\
$\rho^{\rm 7PN}$ &	none &   $13.9527$ 	&  $0.61$  		    &  $1.7\times 10^{-5}$ \\
\hline
$\rho^{\rm 7PN}$ &	$\rho_3$ &   $13.9946$ 	&  $0.58$  		    &  $1.7\times 10^{-5}$ \\
$\rho^{\rm 7PN}$ &	$\rho_3$, $\rho_4^{\rm log}$ &   $14.0015$ 	&  $0.56$  		    &  $1.6\times 10^{-5}$ \\
$\rho^{\rm 7PN}$ &	$\rho_3$, $\rho_4^{\rm log}$, $\rho_5^{\rm log}$ &   $14.00002$ 	&  $0.55$  &  $1.6\times 10^{-5}$ \\
\hline\hline
\end{tabular}
\caption{Numerical determination of the 2PN coefficient $\rho_2$ from the GSF $\rho(x)$ data. Each line describes best-fit results for a particular PN fit model $\rho^{\rm nPN}$ or $\rho^{\rm nPN+}$ [see Eq.\ (\ref{PNmodels}) for notation]. In the upper part of the table we have left all PN parameters of the various models as freely-specifiable fit parameters, whereas in the last 3 lines some of the parameters (specified in the second column) have been fixed at their known analytic values. The third column of the table displays the best-fit value of $\rho_2$ for each model (recall that the analytic value of $\rho_2$ is precisely $\bf 14$), and the fourth column shows the corresponding value of the minimized $\chi^2$ divided by the number of Degrees of Freedom. In the last column we indicate the magnitude of the largest absolute difference 
$\vert \Delta \rho \vert $ between a data point and the value predicted for that point from the best-fit model.
}
\label{Table:rho2}
\end{table}

There are two important caveats one must bare in mind when analyzing the data in Table \ref{Table:rho2}. First, even if the GSF data were {\em exact} (i.e., containing no numerical error), still a fit to a high-order PN-like model would not be expected to produce the exact analytic values of the PN parameters ($\rho_2=14$, etc.) arising from the systematic, mathematically well-defined PN expansion, because the PN series represents an asymptotic expansion while the GSF data includes strong-field information. (A corollary is that the ``best fit'' PN-like model is not at all guaranteed to coincide with the actual, mathematical PN expansion.) The second caveat is that, even if the PN model gave a precise description of $\rho$ for all $x$, still the sum of squares of weighted differences between the GSF data and the PN model would not necessarily follow a $\chi^2$ distribution with mean=DoF, since our numerical data are unlikely to represent independent sample points drawn randomly from a Gaussian distribution. [Our estimated numerical errors, which we treat as random statistical errors for the present exercise, may well have a dominant systematic component (particularly from residual non-stationarity); furthermore, the numerical errors for different $x$ may well be correlated.] Nonetheless, for lack of better options we invoke here the standard $\chi^2$ test as a rough measure of goodness for our fits, and accompany this with $L^{\infty}$-norm information for a fuller picture.

With the above caveats in mind, let us inspect the data in Table \ref{Table:rho2}. We make the following observations. 
(i) The fitted value of $\rho_2$ matches the theoretical PN prediction to within $\lesssim 0.4\%$, so long as the fit model includes all terms through 4PN+ at least. 
(ii) Anecdotically, this agreement becomes extremely good if we allow ourselves to fix some (or all) of the higher-order parameters known analytically.
(iii) For models $\rho^{\rm 5PN}$ and beyond the value of $\chi^2$/DoF is of order unity, indicative of a statistically good fit throughout the entire $x$-domain, and also suggesting that our estimates of the statistical error in $\rho$ are quite reasonable. The slightly low values of $\chi^2$/DoF$\sim 0.6$ (instead $\chi^2$/DoF$\sim 1$) probably reflect our somewhat conservative approach to error estimation in our code (see Ref.\ \cite{Barack:2010tm} for details; for instance, we combine the numerical errors from the various multipole modes by simply adding up their absolute magnitudes rather than adding them in quadrature).   
(iv) The agreement between the fitted $\rho_2$ and its analytic value does not seem to improve (or worsen) significantly upon adding fitting terms beyond 4PN+; likewise, the value of both $\chi^2$/DoF and the $L^{\infty}$-norm seem to ``saturate'' beyond 5PN. It appears that higher-order terms have magnitudes which are too small (relative to the numerical error) to affect our $\chi^2$ analysis (cf.\ the next subsection and Fig.\ \ref{fig:SNR}, where we analyze the magnitude of the various PN ``signals'' as compared with the ``noise'' from numerical error). 
(v) Strikingly, the $L^{\infty}$-norm associated with some of the models in Table \ref{Table:rho2} is extremely low. If one sought to model the GSF-computed $\rho(x)$ at accuracy of (say) better than $10^{-4}$, then our 5PN+ model would be perfectly adequate, even in the strong-field regime. Note that $L^{\infty}\sim 10^{-5}$ is comparable with the magnitude of the statistical error in the GSF data, which explains the saturation of the $L^{\infty}$-norm at this level.

Having established that the GSF data are quantitatively consistent with the analytic 2PN 
result $\rho_2^{\rm analytic}=14$, we henceforth {\em fix} $\rho_2=14$, and turn to consider 
to what extent the GSF data confirm the analytically predicted value of the 3PN coefficient:
$\rho_3^{\rm analytic}=122.6274\ldots$. Once again, we fit various PN models to the numerical data (keeping $\rho_2$ fixed) and this time record the best-fit values obtained for $\rho_3$. The results are presented in Table \ref{Table:rho3}. As before, models $\rho^{\rm 4PN+}$ and beyond yield best-fit $\rho_3$ values in a good agreement with the analytic prediction. The level of agreement ($\lesssim 2\%$ difference) is slightly worse than at 2PN; this is expected given the weaker amplitude of the 3PN ``signal'' (cf.\ Fig.\ \ref{fig:SNR} below).  Once again, models $\rho^{\rm 5PN+}$ and beyond yield $\chi^2$/Dof values of order unity and a $L^{\infty}$-norm smaller than $10^{-4}$. Here, too, fixing the known higher-order coefficients improves the agreement with the analytic prediction significantly. 
\begin{table}[Htb]
\begin{tabular}{lllll}
\hline\hline
fit model$\quad$ & fixed params.~$\quad$ & $\rho_3$ (best fit)$\quad$ &  $\chi^2$/DoF $\quad$ & $L^{\infty}$-norm\\
\hline\hline
$\rho^{\rm 3PN}$ & $\rho_2$	&   $97.953$ 	&  $3.7\times 10^5$ &  $8.2\times 10^{-3}$ \\
$\rho^{\rm 4PN}$ & $\rho_2$ &   $106.936$ 	&  $4.9\times 10^4$ &  $1.2\times 10^{-2}$ \\
$\rho^{\rm 4PN+}$& $\rho_2$	&   $122.458$ 	&  $20.5$  			&  $4.4\times 10^{-4}$ \\
$\rho^{\rm 5PN}$ & $\rho_2$	&   $120.962$ 	&  $12.0$  		    &  $3.6\times 10^{-4}$ \\
$\rho^{\rm 5PN+}$& $\rho_2$	&   $124.365$ 	&  $1.04$  		    &  $7.8\times 10^{-5}$ \\
$\rho^{\rm 6PN}$ & $\rho_2$	&   $122.256$ 	&  $0.57$  		    &  $1.6\times 10^{-5}$ \\
$\rho^{\rm 6PN+}$& $\rho_2$ &   $123.758$ 	&  $0.57$  		    &  $1.6\times 10^{-5}$ \\
$\rho^{\rm 7PN}$ & $\rho_2$	&   $120.914$ 	&  $0.58$  		    &  $1.7\times 10^{-5}$ \\
\hline
$\rho^{\rm 7PN}$ & $\rho_2$, $\rho_4^{\rm log}$ &   $122.929$ 	&  $0.56$  		    &  $1.7\times 10^{-5}$ \\
$\rho^{\rm 7PN}$ & $\rho_2$, $\rho_4^{\rm log}$, $\rho_5^{\rm log}$	&   $122.623$ 	&  $0.55$  		    &  $1.6\times 10^{-5}$ \\
\hline\hline
\end{tabular}
\caption{Numerical determination of the 3PN coefficient $\rho_3$ from the GSF $\rho(x)$ data. 
The structure of the table is similar to that of Table \ref{Table:rho2}, but here we present fit results for $\rho_3$ (recall the analytic value of $\rho_3$ is $\bf 122.6274\ldots$).
In all the models presented we have fixed $\rho_2=14$. The last two lines show results from fits where, in addition, we have fixed the higher-order parameter $\rho_4^{\rm log}$, and then both $\rho_4^{\rm log}$ and $\rho_5^{\rm log}$, at their known analytic values.
}
\label{Table:rho3}
\end{table}

We next apply the same procedure for each of the logarithmic coefficients $\rho_4^{\rm log}$ and $\rho_5^{\rm log}$ in turn: For the former we set both coefficients $\rho_2$ and $\rho_3$ to their exact analytic values, and for the latter we also fix $\rho_4^{\rm log}$. The results are presented in Tables \ref{Table:rho4l} and \ref{Table:rho5l}, respectively. In the case of $\rho_4^{\rm log}$ the agreement with the analytic value is much less impressive than at 2PN and 3PN. Even though models $\rho^{\rm 5PN+}$ and beyond again show $\chi^2$/Dof values of order unity, the resulting best-fit
values for $\rho_4^{\rm log}$ vary within as much as $\sim 35\%$ of the analytic prediction, and the inclusion of additional PN terms does not seem to lead to any convergence of these values. (Note, however, the rather excellent agreement when fixing the higher-order logarithmic coefficient $\rho_5^{\rm log}$.) The agreement is even less striking in the case of $\rho_5^{\rm log}$ (Table \ref{Table:rho5l}), where the best-fit values for models with $\chi^2$/DoF of order unity vary within $\sim 50\%$ of the analytic prediction (but note the rather good---perhaps ``incidental''---agreement for $\rho^{\rm 7PN}$). 
\begin{table}[Htb]
\begin{tabular}{lllll}
\hline\hline
fit model$\quad$ & fixed params.~$\quad$ & $\rho_4^{\rm log}$ (best fit)$\quad$ &  $\chi^2$/DoF $\quad$ & $L^{\infty}$-norm\\
\hline\hline
$\rho^{\rm 4PN+}$& $\rho_2$, $\rho_3$ &   $177.667$ 	&  $50.04$  		&  $6.9\times 10^{-4}$ \\
$\rho^{\rm 5PN}$ & $\rho_2$, $\rho_3$ &   $178.371$ 	&  $51.39$  		&  $7.4\times 10^{-4}$ \\
$\rho^{\rm 5PN+}$& $\rho_2$, $\rho_3$ &   $226.247$ 	&  $1.54$  		    &  $3.2\times 10^{-4}$ \\
$\rho^{\rm 6PN}$ & $\rho_2$, $\rho_3$ &   $169.686$ 	&  $0.56$  		    &  $2.0\times 10^{-5}$ \\
$\rho^{\rm 6PN+}$& $\rho_2$, $\rho_3$ &   $194.132$ 	&  $0.56$  		    &  $1.6\times 10^{-5}$ \\
$\rho^{\rm 7PN}$ & $\rho_2$, $\rho_3$ &   $108.606$ 	&  $0.56$  		    &  $1.7\times 10^{-5}$ \\
\hline
$\rho^{\rm 7PN}$ & $\rho_2$, $\rho_3$, $\rho_5^{\rm log}$ &   $168.474$ 	&  $0.55$   &  $1.6\times 10^{-5}$ \\
\hline\hline
\end{tabular}
\caption{Numerical determination of the 4PN logarithmic coefficient $\rho_4^{\rm log}$ 
from the GSF $\rho(x)$ data. The structure of the table is similar to that of Tables \ref{Table:rho2} and \ref{Table:rho3}, but here we present fit results for $\rho_4^{\rm log}$ (recall the analytic value of $\rho_4^{\rm log}$ is $\bf 167.4666\ldots$). In all cases we have fixed $\rho_2$ and $\rho_3$ at their known analytic values. The last line shows results from a fit where, in addition, we have fixed the higher-order parameter $\rho_5^{\rm log}$. 
}
\label{Table:rho4l}
\end{table}
\begin{table}[Htb]
\begin{tabular}{lllll}
\hline\hline
fit model$\quad$ & fixed params.~$\quad$ & $\rho_5^{\rm log}$ (best fit)$\quad$ &  $\chi^2$/DoF $\quad$ & $L^{\infty}$-norm\\
\hline\hline
$\rho^{\rm 5PN+}$& $\rho_2$, $\rho_3$, $\rho_4^{\rm log}$ &   $-106.14$ 	&  $74.33$  		&  $7.9\times 10^{-4}$ \\
$\rho^{\rm 6PN}$ & $\rho_2$, $\rho_3$, $\rho_4^{\rm log}$ &   $-823.96$ 	&  $0.54$  		    &  $1.6\times 10^{-5}$ \\
$\rho^{\rm 6PN+}$& $\rho_2$, $\rho_3$, $\rho_4^{\rm log}$ &   $-849.15$ 	&  $0.57$  		    &  $1.8\times 10^{-5}$ \\
$\rho^{\rm 7PN}$ & $\rho_2$, $\rho_3$, $\rho_4^{\rm log}$ &   $-1738.65$ 	&  $0.55$  		    &  $1.6\times 10^{-5}$ \\
\hline\hline
\end{tabular}
\caption{Numerical determination of the 5PN logarithmic coefficient $\rho_5^{\rm log}$ 
from the GSF $\rho(x)$ data. The structure of the table is similar to that of Tables \ref{Table:rho2}, \ref{Table:rho3} and \ref{Table:rho4l}, but here we present fit results for $\rho_5^{\rm log}$ (recall the analytic value of $\rho_5^{\rm log}$ is $\bf -1619.4\ldots$). In all cases we have fixed $\rho_2$, $\rho_3$ and $\rho_4^{\rm log}$ at their known analytic values.
}
\label{Table:rho5l}
\end{table}

It is of no surprise that the higher-order PN coefficients are less well determined than the lower-order ones. We expect higher-order PN terms to have lower relative magnitudes (or ``signal''), especially at small $x$, and we heuristically expect the statistical errors in determining each of these terms to be roughly inversely proportional to its typical ``signal-to-noise" ratio (with the ``noise'' here being provided by the statistical numerical error in $\rho$). In addition, in the 4PN and 5PN cases, the signals from the logarithmic terms are strongly correlated with the signals from the unknown 4PN and 5PN `constant' terms, which  further reduces the extraction accuracy of the logarithmic coefficients.

It is instructive to examine the magnitudes of the various PN signals relative to the numerical noise (as was also done in Ref.\ \cite{Blanchet:2010zd}), and we do so in the next subsection.

\subsection{``Signal-to-noise'' analysis}\label{Sec:SNR}

\begin{figure}[htb]
\includegraphics[width=12cm]{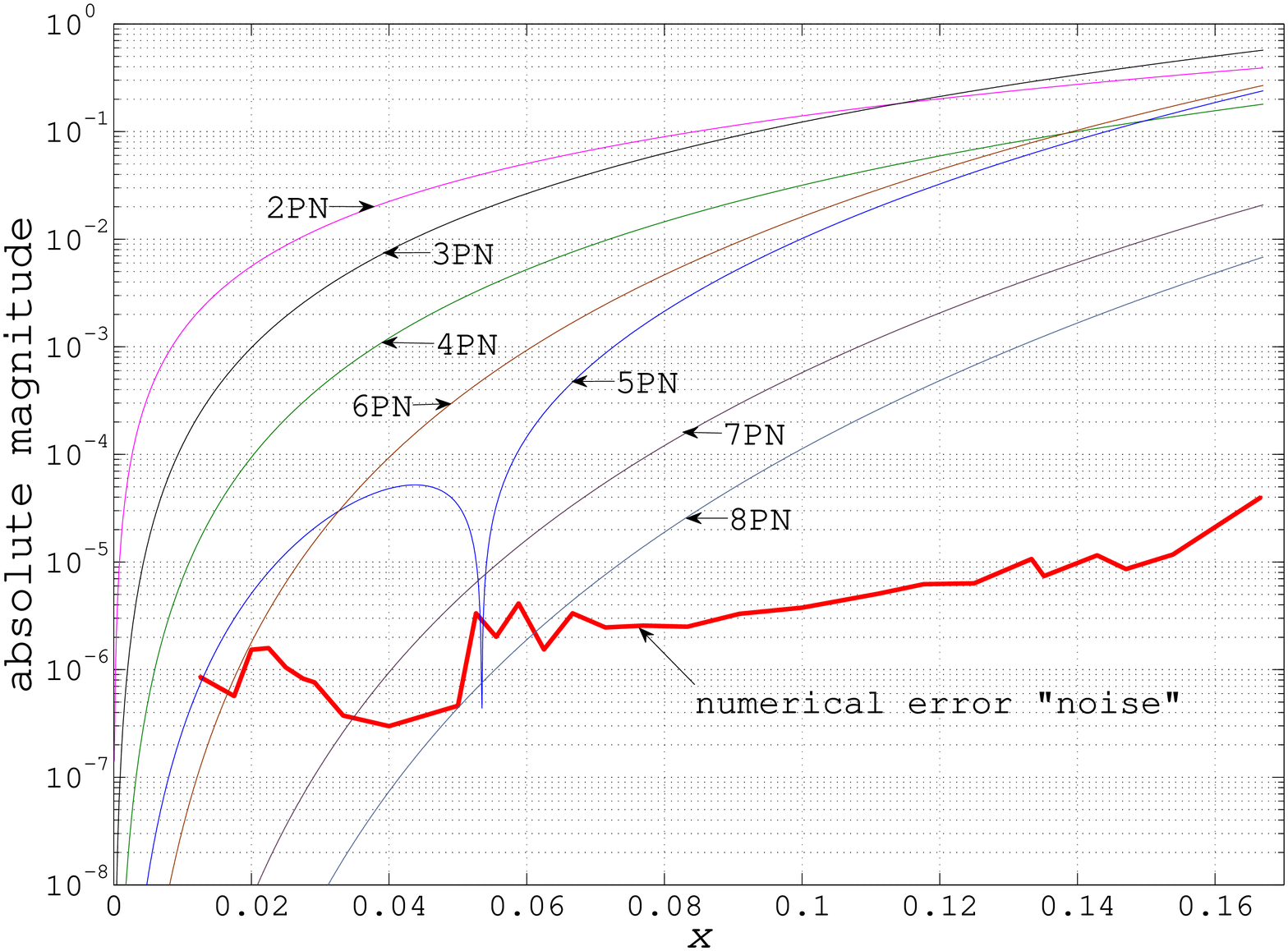}
\caption{``Signal'' amplitudes from various PN contributions to $\rho(x)$ (thin lines), compared with the ``noise'' amplitude from numerical error (thick red line). For this plot, all unknown PN coefficients through 8PN were crudely estimated by fitting the numerical GSF data to a tentative 8PN model. Each of the signal lines displays the amplitude of the {\em total} contribution from a particular PN order, which at 4PN through 6PN also includes a logarithmic term of the form $\rho_n x^n\ln x$ (at 7PN and 8PN we crudely absorb all possible logarithmic running in an effective term of the form $\rho_n x^n$). Note the logarithmic scale of the vertical axis. The total 5PN contribution changes its sign around $x=0.0535$, where the (negative) $\rho_5^{\rm c}x^5$ term conspires to cancel out the (positive) $\rho_5^{\rm log}x^5\ln x$ term. (Note that in this figure the labels `nPN' indicate the contribution of an {\em individual} PN order, unlike elsewhere in the text where nPN stands for a partial PN sum.) 
}
\label{fig:SNR}
\end{figure}

We are interested here in a rough estimation of the magnitudes of the various PN contributions to $\rho(x)$, in comparison with the numerical error in $\rho$. This requires knowledge of some of the yet-unknown high-order PN coefficients. In the next section we attempt to determine some of these coefficients through a systematic analysis. For our present purpose, however, we shall content ourselves with a cruder approach, whereby we fit all unknown PN coefficients simultaneously using a simple, tentative high-PN model. Specifically, we fit the numerical data to an 8PN model in which all known parameters ($\rho_2$, $\rho_3$, $\rho_4^{\rm log}$ and $\rho_5^{\rm log}$) are pre-fixed at their analytic values, and where $\rho_4^{\rm c}\ldots \rho_8^{\rm c}$, as well as $\rho_6^{\rm log}$, are left as fitting parameters. Since the form of the logarithmic dependence at high PN order is not yet clear [there may well occur $\propto (\ln x)^2$ terms, for example] we choose to crudely absorb all possible 7PN terms in a term of the standard form $\rho_7^{\rm c} x^7$, and similarly for 8PN. A least-squares fit then yields the following values (which one should consider merely indicative):  $\rho_4^{\rm c}=68.48$, $\rho_5^{\rm c}=-4742.81$, $\rho_6^{\rm c}=-771.41$, $\rho_6^{\rm log}=-7349.30$, $\rho_7^{\rm c}=5757.52$, and $\rho_8^{\rm c}=11278.11$. We use these values to deduce the amplitudes of the PN contributions through 8PN, as functions of $x$. We show these amplitudes in Figure \ref{fig:SNR}, along with the amplitude of the ``noise'' from numerical error. 

The accuracy $\Delta\rho_n$ with which we can extract the value of a given PN coefficient $\rho_n$ (assuming all lower-order coefficients are known) is roughly inversely proportional to the typical signal-to-noise ratio (SNR) associated with the corresponding (individual) nPN signal. Inspection of Figure \ref{fig:SNR} reveals that the SNRs of the 2PN and 3PN signals are roughly equal in the strong field ($x\gtrsim 0.1$), but in the weak field the 2PN SNR increases gradually with respect to the 3PN one, up to about a relative factor 10 at $x=1/80$. This is consistent with our finding that $\Delta\rho_2$ is about 5 times smaller than $\Delta\rho_3$. 

The 3PN SNR, in turn, is always larger than the 4PN SNR, by an amount varying from a factor $\sim 3$ near the ISCO to a factor $\sim 10$ in the weak field. The extraction error $\Delta\rho_4^{\rm log}$ was found to be $\sim15$--$20$ times larger than $\Delta\rho_3$, which is a bit more than one might expect based on the simple `1/SNR' scaling argument. However, here one should recall that the problem of fitting $\rho_4^{\rm log}$ also involved fitting the unknown parameter $\rho_4^{\rm c}$, which necessarily has the effect of further increasing the error $\Delta\rho_4^{\rm log}$. Hence, our ``empirical'' level of uncertainty in $\rho_4^{\rm log}$ is not unreasonable.

The case of the 5PN signal is more involved. As manifest in Figure \ref{fig:SNR}, the 5PN signal is strongly suppressed, and it entirely vanishes around $x=0.0535$. This is a result of a cancellation between the negative term   $\rho_5^{\rm c} x^5\cong-4742.81x^5$ and the positive term $\rho_5^{\rm log} x^5\ln x\cong -1619.428571 x^5\ln x$ (recall $\ln x<0$ in the relevant domain). Inspection of the separate SNRs from the two 5PN contributions (not shown in the plot) reveals that they are each about equal to the 4PN SNR in the strong field ($x\gtrsim 0.1$), and gradually decrease in the weak field to about $1/10$ of the 4PN SNR at $x=1/80$. However, our simple `1/SNR' scaling argument clearly does not apply here, since the two 5PN signals are manifestly strongly correlated between themselves, and, moreover, the effective noise level for the 5PN signals has a large contribution from the poorly resolved coefficient $\rho_4^{\rm c}$. These complications suggest that $\rho_5^{\rm log}$ should be rather poorly resolved using our current GSF data, as we indeed concluded above through experiment.\footnote {One might attempt to make a more quantitative prediction for $\Delta\rho_5^{\rm log}$ through analysis of the variance-covariance matrix defined on the signal parameter space. Here, however, we content ourselves with a more heuristic discussion.}

The information in Figure \ref{fig:SNR} serves not only to check that our fitting accuracies for the known PN parameter make sense, but also (more importantly) to establish an expectation as to which of the yet-unknown PN coefficients we might hope to resolve using our current GSF data. It is immediately clear, for example, that we are not likely to be able to extract the 7PN and 8PN coefficients at any level of confidence, since the crucial weak-field portion of their signal is deeply buried in the numerical noise (and, furthermore, the strong-field portions of the 7PN and 8PN signals are likely to be strongly correlated, as is visually evident from the near-proportionality of the two signals in Figure \ref{fig:SNR}).  

The situation with the 6PN signal is somewhat better: We have statistically-significant data available for $x\gtrsim 0.02$, so one might hope to be able to extract some information about the 6PN coefficients. However, since the 4PN and 5PN parameters are not known analytically, the extraction errors from these parameters will effectively add to the noise level relevant for the 6PN signal. Furthermore, since at present the 6PN logarithmic dependence is not known analytically, it might prove difficult to disentangle (with statistical confidence) the $\rho_6^{\rm c} x^6$ term from any possible 6PN logarithmic terms. 

As for the 4PN and 5PN terms: Here the signals lie above the numerical noise across the entire $x$ domain of the numerical data, and, in addition, the logarithmic running terms are known analytically. This suggests that our numerical data is sufficiently accurate to allow fitting the values of $\rho_4^{\rm c}$ and $\rho_5^{\rm c}$. Here too, however, we note that the effective noise level for the 5PN term will be increased by the statistical extraction error from the 4PN term. 

The above is, of course, only a heuristic discussion based on the tentative signal amplitudes shown in Figure \ref{fig:SNR}. In the next section we carry out a more systematic analysis, aimed at extracting the values of (and determining the error bars on) as many PN parameters as is possible given the numerical data.

\section{Determination of unknown EOB/PN parameters}

The main potential payoff from a GSF--EOB synergy lies in the determination of the {\em strong-field} behavior of some $O(q)$ functions relevant to the dynamics of binary systems. In this section, however, we remain for the time being within the context of PN theory, and attempt to use the GSF data to determine some of the unknown higher-order weak-field parameters entering the PN expansion of $\rho(x)$. In the following section we will then turn to explore the strong-field information contained in our GSF data.

\subsection{Numerical determination of yet-unknown, higher-order PN expansion parameters of $\rho(x)$} 

 The heuristic discussion in the previous section suggested that we might be able to determine $\rho_4^{\rm c}$ and possibly $\rho_5^{\rm c}$, while the ability to determine the 6PN parameters remains less certain. The tentative signal amplitudes in Figure \ref{fig:SNR} also suggest that in fitting a PN model to the GSF data we should include all terms down to 7PN, or perhaps 8PN (otherwise, the remaining  ``unmodelled'' piece in the numerical data would significantly affect the estimates
 of the lower PN parameters). 
As a final preliminary note, we observe (extrapolating the upper two curves in Figure \ref{fig:Delta4} ``by eye'' to $x=0$) that the value of $\rho_4^{\rm c}$ is expected to fall in the range $50<\rho_4^{\rm c}< 100$. 

Our analysis proceeds as follows. Fixing all known parameters ($\rho_2$, $\rho_3$, $\rho_4^{\rm log}$, $\rho_5^{\rm log}$) at their analytic values, we consider a set of PN models from $\rho^{5{\rm PN}+}$ through to $\rho^{8{\rm PN}}$. At 7PN and 8PN we attempt a variety of 7PN logarithmic dependences, including $x^7\ln x$ and $x^7(\ln x)^2$ (each in separate or combined forms). We fit each model to the GSF data and record the best-fit values of $\rho_4^{\rm c}$, $\rho_5^{\rm c}$, $\rho_6^{\rm c}$ and $\rho_6^{\rm log}$, along with the values of $\chi^2$/DoF and the $L^{\infty}$-norm. The results are displayed in Table \ref{Table:rho456}.
\begin{table}[Htb]
\begin{tabular}{lcccccc}
\hline\hline
fit model  &$\rho_4^{\rm c}$ &$\rho_5^{\rm c}$ &$\rho_6^{\rm c}$ & $\rho_6^{\rm log}$ & $\chi^2$/DoF  & $L^{\infty}$-norm \\
\hline\hline
$\rho^{\rm 5PN+}$           &  
$-47.0679$  &  $-1850.32$  	& 	{--}    & {--}     &  $2.2\times 10^5$ & $1.0\times 10^{-2}$ \\
$\rho^{\rm 6PN}$            &  
$43.1869$   &  $-3544.91$  	& $7428$    & {--}     &  $89.84$          & $7.6\times 10^{-4}$ \\
$\rho^{\rm 6PN+}$           &  
$62.9429$   &  $-4388.12$  	& $4346$    & $-4124$  &  $1.41$     & $9.8\times 10^{-5}$ \\
$\rho^{\rm 7PN}$            &  
$69.3494$   &  $-4824.12$  	& $-3920$   & $-8682$  & $0.52$      & $1.6\times 10^{-5}$ \\
$\rho^{\rm 7PN+}$           & 
$68.3530$   &  $-4719.34$  	& $2621$    & $-6395$  &  $0.55$     & $1.6\times 10^{-5}$ \\
$(\rho^{\rm 7PN})^{\dagger}$&  
$65.6983$   &  $-4444.37$  	& $19456$   & $-479$   &  $0.53$     & $1.7\times 10^{-5}$ \\
$(\rho^{\rm 7PN})^\ddagger$ &  
$69.6710$   &  $-4826.95$  	& $6126$    & $-6646$  &  $0.53$     & $1.6\times 10^{-5}$ \\
$(\rho^{\rm 7PN+})^\sharp$  &  
$75.6533$  	& $-5933.41$    & $-217404$ & $-62548$ & $0.56$      & $1.7\times 10^{-5}$ \\
$\rho^{\rm 8PN}$            &  
$73.2926$   &  $-5451.35$   & $-84429$  & $-31845$ &  $0.56$     & $1.7\times 10^{-5}$ \\
($\rho^{\rm 8PN})^\dagger$  &  
$69.0076$   &  $-4812.02$  	& $-7820$   & $-9484$  &  $0.54$     & $1.6\times 10^{-5}$ \\
($\rho^{\rm 8PN})^\ddagger$ &  
$68.3559$   &  $-4724.35$  	& $2380$    & $-6546$  &  $0.56$     & $1.7\times 10^{-5}$ \\
($\rho^{\rm 8PN})^\S$ &  
$68.4819$   &  $-4742.81$  	& $-771$    & $-7349$  &  $0.54$     & $1.6\times 10^{-5}$ \\
\hline\hline
\multicolumn{7}{l}
{$^\dagger$replacing the term $\rho_7^{\rm c} x^7$ with $\rho_7^{\rm log}x^7 \ln x$}\\
\multicolumn{7}{l}
{$^\ddagger$replacing the term $\rho_7^{\rm c} x^7$ with $\tilde\rho_7^{\rm log}x^7 (\ln x)^2$}\\
\multicolumn{7}{l}
{$^\sharp$adding a term $\tilde\rho_7^{\rm log}x^7 (\ln x)^2$}\\
\multicolumn{7}{l}
{$^\S$omitting the term $\rho_7^{\rm log}x^7 \ln x$}
\end{tabular}
\caption{Numerical determination of yet-unknown, higher-order PN expansion parameters of $\rho(x)$ from GSF data. Each line in the table corresponds to a particular PN fit model as indicated in the first column [referring to Eq.\ (\ref{PNmodels}) for notation]. In all models we have fixed $\rho_2$, $\rho_3$, $\rho_4^{\rm log}$ and $\rho_5^{\rm log}$ in accordance with their known analytic values [Eqs.\ (\ref{rho23}) and (\ref{rho45log})]. For each model we give the best-bit values of $\rho_4^{\rm c}$, $\rho_5^{\rm c}$, $\rho_6^{\rm c}$ and $\rho_6^{\rm log}$, and the corresponding values of $\chi^2$/DoF and the $L^{\infty}$-norm. 
Note the last model in the table is the tentative one considered in Subsec.\ \ref{Sec:SNR}. 
}
\label{Table:rho456}
\end{table}

We observe that, in accordance with our expectation, the goodness of the fit (as measured by $\chi^2$/DoF and the $L^{\infty}$-norm) ``saturates'' at the 7PN level; lower-order PN models do not fit the numerical data as well, and the fit does not seem to improve when including 8PN terms. We therefore focus our attention on the 7PN and 8PN models in Table \ref{Table:rho456}. These models predict $\rho_4^{\rm c}$ values between $65.6983$ and $75.6533$. We consider this interval a measure of the uncertainty, $\Delta\rho_4^{\rm c}$, in our determination of $\rho_4^{\rm c}$. We note, however, that many of the $\rho_4^{\rm c}$ values in the table are clustered around 68 or 69. We shall take $\rho_4^{\rm c}=69$ as our ``best guess'' value, and as a rough error margin take the asymmetric range between $65$ and $76$. The values predicted for $\rho_5^{\rm c}$ vary between $-4444.37$ and $-5933.41$, with many of the values clustered around $-4700$ or $-4800$. We shall estimate $\rho_5^{\rm c}=-4800$, with asymmetric error range between $-6000$ and $-4400$. The values for $\rho_6^{\rm c}$ are clearly dominated by random noise, and provide us with no meaningful information. The best fit values $\rho_6^{\rm log}$ are likewise very ``noisy'', although is seems safe to conclude that the actual value of  $\rho_6^{\rm log}$ is negative (and perhaps of order a few thousands). As expected, the values of the various 7PN and 8PN coefficients (not shown in Table \ref{Table:rho456}) are entirely dominated by noise and cannot be determined.

In summary, the information included in the currently available GSF data allows us to conclude
\begin{equation}\label{calibrated}
\rho_4^{\rm c}=69^{+7}_{-4}, \quad\quad \rho_5^{\rm c}=-4800^{+400}_{-1200}, \quad\quad \rho_6^{\rm log}<0.
\end{equation}
We note that the rather large uncertainties in $\rho_4^{\rm c}$ and $\rho_5^{\rm c}$ are in fact similar in magnitude to the ones obtained in the previous section for (correspondingly) $\rho_4^{\rm log}$ and $\rho_5^{\log}$---cf.\ Tables \ref{Table:rho4l} and \ref{Table:rho5l}. Somewhat disappointingly, the accuracy of our numerical data does not allow us to set tighter constraints on $\rho_4^{\rm c}$ and $\rho_5^{\rm c}$, nor does it give us reliable access to the values of $\rho_6^{\rm c}$ and $\rho_6^{\rm log}$.

\subsection{Implications of $\rho_4^{\rm c}$ and $\rho_5^{\rm c}$ for EOB theory}\label{subsec:implications}

What can we learn from the above estimates of $\rho_4^{\rm c}$ and $\rho_5^{\rm c}$ about the values of the yet-undetermined $O(q)$ (logarithmically running) EOB parameters 
$a_5(\ln x),a_6(\ln x),\ldots$ and $\bar d_4(\ln x),\bar d_5(\ln x),\ldots$?
To discuss this issue we need to come back to the exact EOB relation between the GSF-determinable
function $\rho(x)$ and the basic functions $a(x)$ and $\bar d(x)$ of the EOB formalism, i.e., to
Eqs.\ (\ref{rhoEOB})--(\ref{rhoEOBpieces}). Above, we considered the first two terms in 
the PN expansion of these equations, corresponding to the 2PN and 3PN levels. If we now consider
the higher-order terms in the PN expansion of these equations, and if we separate them according to
whether they contain $\ln x$ or not, 
we find that $\rho_4^{\rm c}$ and $\rho_5^{\rm c}$ are
related to the coefficients entering the PN expansions of the $a(x)$ and $\bar d(x)$ functions 
[setting $a_5(\ln x)=a_5^{\rm c} + a_5^{\rm log} \ln x$, etc.] via
\begin{eqnarray}
\rho_4^{\rm c} &=& -\frac{27}{4}-7a_4+10 a_5^{\rm c}-6 \bar d_3+\bar d_4^{\rm c} +\frac{9}{2}a_5^{\rm log}, \label{rho4ad} \\
\rho_5^{\rm c} &=& -\frac{675}{32}-14 a_5^{\rm c}+15 a_6^{\rm c}-6\bar d_4^{\rm c}+\bar d_5^{\rm c}-8 a_5^{\rm log} +\frac{11}{2} a_6^{\rm log}. \label{rho5ad}
\end{eqnarray}
Substituting from Eqs.\ (\ref{a4}), (\ref{d3}) and (\ref{calibrated}) we then obtain the constraints 
\begin{eqnarray}\label{constraint1}
10 a_5^{\rm c}+\bar d_4^{\rm c}+\frac{9}{2}a_5^{\rm log} & \simeq &  518.6^{+7}_{-4}, \\
\label{constraint2}
14 a_5^{\rm c}+6\bar d_4^{\rm c}-15 a_6^{\rm c}-\bar d_5^{\rm c}+8 a_5^{\rm log} -\frac{11}{2} a_6^{\rm log} &\simeq & 4779^{-400}_{+1200}.
\end{eqnarray}
 We note the fortunate fact that the large relative uncertainty in our fit for $\rho_4^{\rm c}$ manifests itself only with a modest fractional error on the right-hand side of the constraint equation (\ref{constraint1}) [this is because the term $\rho_4^{\rm c}\sim 69$ in  Eq.\ (\ref{rho4ad}) happens to be quite a bit smaller than the known part $7a_4+6\bar d_3\simeq 442.815$ in that equation]. 
 On the other hand, the fractional uncertainty on the right-hand side of the constraint equation (\ref{constraint2}) remains rather large.

As noted in Ref.\ \cite{Damour:2009sm}, knowledge of the PN expansion coefficients of $\rho(x)$ does not on its own allow us to constrain separately the PN expansion coefficients of the
EOB functions $a(x)$ and $\bar d(x)$, but only combinations thereof. Another subtlety which enters
our constraints (\ref{constraint1}), (\ref{constraint2}) is that their left-hand sides involve
a combination of  non-logarithmic and logarithmic PN coefficients. Our GSF results do not either
give us a direct access to the separate values of $a_5^{\rm log}$ and $a_6^{\rm log}$, but only
to the specific combinations of $a_5^{\rm log}$, $\bar d_4^{\rm log}$, $a_6^{\rm log}$ and $\bar d_5^{\rm log}$
that enter $\rho_4^{\rm log}$ and $\rho_5^{\rm log}$, namely
\begin{eqnarray}
\rho_4^{\rm log} &=& +10 a_5^{\rm log}+\bar d_4^{\rm log}, \label{rho4log} \\
\rho_5^{\rm log} &=& -14 a_5^{\rm log}+15 a_6^{\rm log}-6\bar d_4^{\rm log}+\bar d_5^{\rm log}. \label{rho5log} 
\end{eqnarray}

Let us note in passing that Ref.\ \cite{Damour:2009sm} has discussed an approximate way of
estimating some `effective' values of the coefficients $a_5$ and $a_6$, roughly corresponding
to averages of the logarithmically running parameters $a_5(\ln u)$ and $a_6(\ln u)$ over
an interval of $u$ around the ISCO. With the present limited information that we can derive
from our GSF data, we cannot meaningfully combine these approximate estimates with our
results (\ref{constraint1}), (\ref{constraint2}).
Ref.\ \cite{Damour:2009sm} went on to suggest gauge-invariant quantities other than $\rho$ (namely, the ``whirl'' frequency and angular momentum in a zero-binding zoom-whirl orbit), which give a handle on the strong-field behavior of the $a$ function {\em without} involving the $\bar d$ function. 
We shall come back below to possible ways of combining the knowledge of such additional quantities
with the results that can be derived from our present GSF data.

\section{On the determination of the global strong-field behavior of various $O(\nu)$ EOB functions}

So far we have used our $\rho(x)$ data to (i) test the GSF (and the PN) calculation(s) and (ii) constrain some of the unknown higher-order PN parameters entering the weak-field limit of  EOB theory. 
However, as already mentioned above, the main potential payoff from a synergy between GSF and EOB formalisms is the determination of the strong-field behavior of some $O(\nu)$ functions relevant to the dynamics of the binary system. Indeed, the EOB formalism has shown its ability at accurately describing
all phases of the merger process of comparable-mass binaries, from the early inspiral to the final ringdown \cite{Damour:2009kr,Buonanno:2009qa}, in terms of a few basic functions, and notably
the functions $A(u,\nu)$ and $\bar D(u,\nu)$ whose $O(\nu)$ parts are the functions $a(u)$ $\bar d(u)$.
Our numerical results for $\rho(x)$ therefore give us, through the relations (\ref{rhoEOB})--(\ref{rhoEOBpieces}), a first-ever access to the {\em strong-field behavior} of (a combination of) functions entering the description of the binary's conservative dynamics. As such, our results have the potential to inform the development of the EOB formalism,  which is of great interest and timeliness in the context of the vigorous effort to model gravitational-wave sources for existing and planned detector projects. In this context, it would be desirable to replace our sample of tabulated values of the function $\rho(x)$, Table \ref{Table:data}, by some analytical representation which faithfully matches our numerical results throughout the entire domain $0< x\leq 1/6$, and which is likely to remain adequate even for larger values of $x$.

\subsection{Accuracy threshold on the global representation of $\rho(x)$}

Before discussing various ways which might be used for obtaining such global analytical
representations of our strong-field data, one should start by assessing the accuracy requirements
that such global representations must satisfy. In the following we suggest a way of quantifying the desired accuracy standard for $\rho(x)$.

The currently most accurate version
of the EOB formalism \cite{Damour:2009kr} has found, especially for the equal-mass case, 
an excellent fit (within the NR error bar) between the EOB waveform and the NR 
Caltech-Cornell waveform \cite{Scheel:2008rj} all over a certain thin `banana-like' region
in the plane of the two effective parameters $a_5^{\rm effective}$, $a_6^{\rm effective}$
which are used to parametrize the form of the global, Pad\'e-resummed function 
$A_P(u,\nu;a_5^{\rm effective},a_6^{\rm effective})$. Comparing the variation of $A_P$ 
along the center of the good-fit region  (for
$\nu=1/4$) with the variation of an $A$ function of the type considered here [i.e.,
$A(u,\nu)=1-2u+\nu a(u) +O(\nu^2)$, where $a(u)$ is
allowed to change while the $O(\nu^2)$ terms are kept fixed], allows us to relate the
variation of $A_P(u)$ to the variation of $\nu a(u)$ (with $\nu=1/4$). Dividing by $\nu=1/4$,
one can then use this variation of the function  $ a(u)$ to infer the corresponding variation of 
the function $ \rho(u)$ using the relations (\ref{rhoEOB})--(\ref{rhoEOBpieces}). (In these mathematical
expressions one can rename the variable $u$ as $x$.) Using this procedure, one finds that
the resulting variation of the function $\rho(x)$, as one moves along the center of the good-fit region,
stays globally quite small all over the strong-field interval $0\leq x\leq 1/3$ explored
by the EOB evolution. In particular, restricting to the interval $0\leq x\leq 1/6$ currently accessible to GSF methods, one finds that the variation of $\rho(1/6)$ which is compatible with the
current NR/EOB agreement is between $-0.0028$ (for the lefmost part of the banana-like
good-fit region) and $+0.0028$ (for its rightmost part). [Here, we use as a reference point
for evaluating the variation the value of $\rho(x)$ corresponding to the `intersection values'
($a_5^{\rm effective}=-22.3, a_6^{\rm effective}=+252)$ selected in Ref.\ \cite{Damour:2009sm}.]
When extending the strong-field interval as far as the `light ring' \footnote{In the EOB formalism
the functions $A(u)$, $\bar D(u)$, etc.\ are supposed to have no singularity at least up
to values of their arguments of order $u\sim 1/2 +O(\nu)$ corresponding to the EOB
formal analog of the `horizon'. But a crucial role is played by their behavior up to
the EOB analog of the light-ring, i.e., for $x=1/3 +O(\nu)$. Note that Eqs.\ (\ref{rhoEOB})--(\ref{rhoEOBpieces}) suggest that the function $\rho(x)$ will be singular
at $x=1/3$ [because of the $\rho_E(x)$ contribution]. However, our discussion is meaningful for
the `corrected' function $\tilde \rho(x) \equiv \rho(x)-\rho_E(x)= \rho_a(x)+\rho_d(x)$,
which is indeed predicted by the EOB formalism to be regular up to $u\sim 1/2 +O(\nu)$.}, i.e., $x=1/3$,
one finds that the `allowed' variation of $\rho(1/3)$ is somewhat larger, and of order $\pm 0.01$
(and even more in the leftmost region).

Summarizing: The minimal accuracy requirements that one should impose (at present) on a global
representation of the function $\rho(x)$ is a maximum abolute difference (i.e.,
an $L^{\infty}$-norm) smaller or equal to $ 2.8\times 10^{-3}$ over the interval $0\leq x\leq 1/6$, and staying smaller than about $10^{-2}$ over the full interval $0\leq x\leq 1/3$ where GSF $\rho$-data might
eventually be available.

\subsection{On the use of actual, PN expansions for representing $\rho(x)$ globally}

One can think of several options for constructing global analytical representations of the
function $\rho(x)$ from our GSF results.

A first option might be to try to represent $\rho(x)$ by a sufficiently large number
of terms of its (actual) PN expansion in powers of $x$ (including all needed logarithms).
However, our Figure \ref{fig:SNR} clearly shows that this is an impossible task.
Even if one limited one's ambition to representing $\rho(x)$ by its PN expansion on the 
interval $0\leq x\leq 1/6$, Figure \ref{fig:SNR} shows that one should first determine the PN
expansion coefficients of $\rho(x)$  beyond the 8PN level, and possibly beyond the 9PN level.
Indeed, if we use the fitted values of $\rho_7$ and $\rho_8$ quoted above
(where, for simplicity, we drop the superscript c) as being indicative of their real values 
(when absorbing the logarithms in the corresponding PN terms), we have 
$\rho_7 x^7 \simeq 0.0206 (6x)^7$, and $\rho_8x^8 \simeq 0.0067 (6x)^8$, which fail to
reach the required level $0.0028$ for $x=1/6$. From the ratio of these two terms 
(D'Alembert criterion) \footnote{The ratio $\rho_8/\rho_7 \simeq 1.96$ suggests a radius of
convergence around $x\simeq 1/2$; this, in turn, suggests that the exact PN series might still
converge when $x \sim 1/3$, but at an extremely slow rate $\sim \sum_n C (2 x)^n$. To be precise
one should discuss here the convergence of the `corrected' function $\tilde \rho(x) \equiv \rho(x)-\rho_E(x)$. Indeed, the piece $\rho_E(x)$ will ultimately limit the convergence
radius of $\rho(x)$ to $1/3$ because of its singular behavior there. However, one can check
that, even at the 9PN level, the contribution of the PN expansion of  $\rho_E(x)$ to that
of  $\rho(x)$ is numerically small compared to $\rho_7$, $\rho_8$ and our estimated $\rho_9$,
and would therefore not affect our conclusion.}
one expects the 9PN term to be of order $0.0022 (6x)^9$, and to barely meet our requirement.
However, it is currently unthinkable to be able to either GSF-numerically compute
all the needed PN coefficients with a decent accuracy \footnote{We recall that the extraction of accurate PN
coefficients  requires accurate GSF data at large binary separations (small $x$), and that obtaining such data is extremely computationally costly for our eccentric-orbit problem.} 
or to derive them analytically. Moreover, even if we had at hand these higher-order PN terms, they would be essentially useless for controlling the value of the function $\rho(x)$ in the strong-field interval $1/6 <x<1/3$
which is crucial for the applicability of the EOB formalism. Indeed, the formal computation
of the 7PN , 8PN and 9PN contributions we just wrote down in the extended interval $1/6 <x<1/3$
shows that the PN expansion completely loses its numerical validity when $x \sim 1/3$
(for instance $\rho_8x^8 \simeq 1.7 (3x)^8$ is formally of order unity when $x \sim 1/3$).
Evidently, all this is an illustration of the fact that the PN series, being a weak-field
expansion, cannot be expected to cover the {\em very-strong-field regime} $1/6 <x<1/3$ corresponding to
radii ranging between the ISCO and the light-ring. [Recall, however, that the Pad\'e-resummed
functions used in the EOB formalism seem adequate to accurately describe this regime.]

\subsection{On the use of effective, PN-like expansions for representing $\rho(x)$ globally}

If the {\em actual} PN expansion of the function $\rho(x)$ (corresponding to the
mathematically exact Taylor-type expansion around $x=0$) is inadequate for defining
a global representation of this function, one might think that some type of {\em effective}
PN-type expansion of $\rho(x)$ might do a better job. Indeed, we have seen above that
by fitting various high-order PN-type models to our numerical GSF data we could represent the
function  $\rho(x)$ in the interval $0\leq x\leq 1/6$ by polynomials (with extra logarithms)
with $L^{\infty}$-norm values as small as $\sim 10^{-5}$. Two of the remarkable aspects of our
`experimental' findings in Sec.\ \ref{comparison} above were: (i) one could obtain 
$L^{\infty}$-norms of order $\sim  {\rm few} \, 10^{-5}$ (i.e., at the level of our statistical numerical error) by using much fewer terms in the PN-like expansion
than expected from our rough convergence analysis in the previous subsection (e.g., 5PN with
logarithms, instead of 9PN from the expected convergence rate); and (ii) one
could (nearly) obtain the required accuracy level $L^{\infty}$-norm $\sim {\rm few} \, 10^{-3}$ by using as low a formal PN accuracy as 3PN (when fitting for $\rho_2$ and $\rho_3$ {\em without} fixing
them by our analytical knowledge). However, these results apply only to the restricted
interval $0\leq x\leq 1/6$, and do not give us any control on the type of PN-like expansions 
needed to globally represent $\rho(x)$, within the accuracy threshold mentioned above, 
on the {\em doubled} interval $0\leq x\leq 1/3$. From Weierstrass' Approximation Theorem 
we know that the (arguably) continuous function $\rho(x)$ [or rather $\tilde \rho(x)$] can be uniformly approximated, on any compact interval, by
some finite polynomial in $x$ to any required accuracy, but we do not know anything
about the order of this polynomial. In addition, if we lose the connection between the
coefficients of this polynomial and the Taylor coefficients of $\rho(x)$ at $x=0$, we lose the advantage of having an advance analytic knowledge of some of these coefficients. Lastly, we should recall that constructing an accurate PN-like model based on a fit to GSF data entails having at hand a dense sample of GSF data points across the entire range $0<x<1/3$. Obtaining such a data set can be extremely computationally expensive.

We conclude from this discussion that any type of PN expansion (actual or effective) is
ill-suited for obtaining a global representation of $\rho(x)$ that meets the rather
strict accuracy requirements discussed above in the full desired interval
$0\leq x\leq 1/3$ (of which only the first half has yet been covered by our current GSF results).

\subsection{Accurate global representations of $\rho(x)$ using multiple-point Pad\'e approximants}

We wish to introduce here a new strategy for defining  sufficiently accurate global 
representations of the strong-field behavior of dynamically useful functions based on 
the combined use of a {\em minimal} amount of strong-field information,
together with the currently available weak-field (i.e., PN) information.
This strategy has some similarity with the strategy used in defining the
EOB basic functions, and notably $A(u,\nu)$, by means of Pad\'e approximants, but
it goes further in its way of incorporating some strong-field information.
Indeed, the Pad\'e approximants used so far in EOB theory have always been
{\em one-point} Pad\'e approximants (i.e., rational functions constrained to
reproduce some given Taylor series around, say, the point $x=0$). By contrast,
the general strategy we propose here consists in using 
{\em multiple-point} Pad\'e approximants, i.e., rational functions constrained to
reproduce {\em several} given Taylor series around several  points, say $x_1=0, x_2, x_3, \ldots$. 
One of the expansion points at which one injects some (PN) information is still the
weak-field limit $x=0$, but the other points $x_2, x_3, \ldots$ will be taken in the
strong-field domain, and will be used to inject a limited, but crucial amount of
strong-field information (obtained, say, by a correspondingly limited number
of GSF computations). Here we shall demonstrate the applicability of our method by considering the function $\rho(x)$ \footnote{It might even be better to apply it to the function $\tilde \rho(x)$, but we shall not bother to
do that here, because we have in mind other applications, which we mention below.}.
However, the basic idea  is clearly very general, and we believe it can provide a powerful tool for advancing AR technology. 

\begin{figure}[Htb]
\includegraphics[width=11cm]{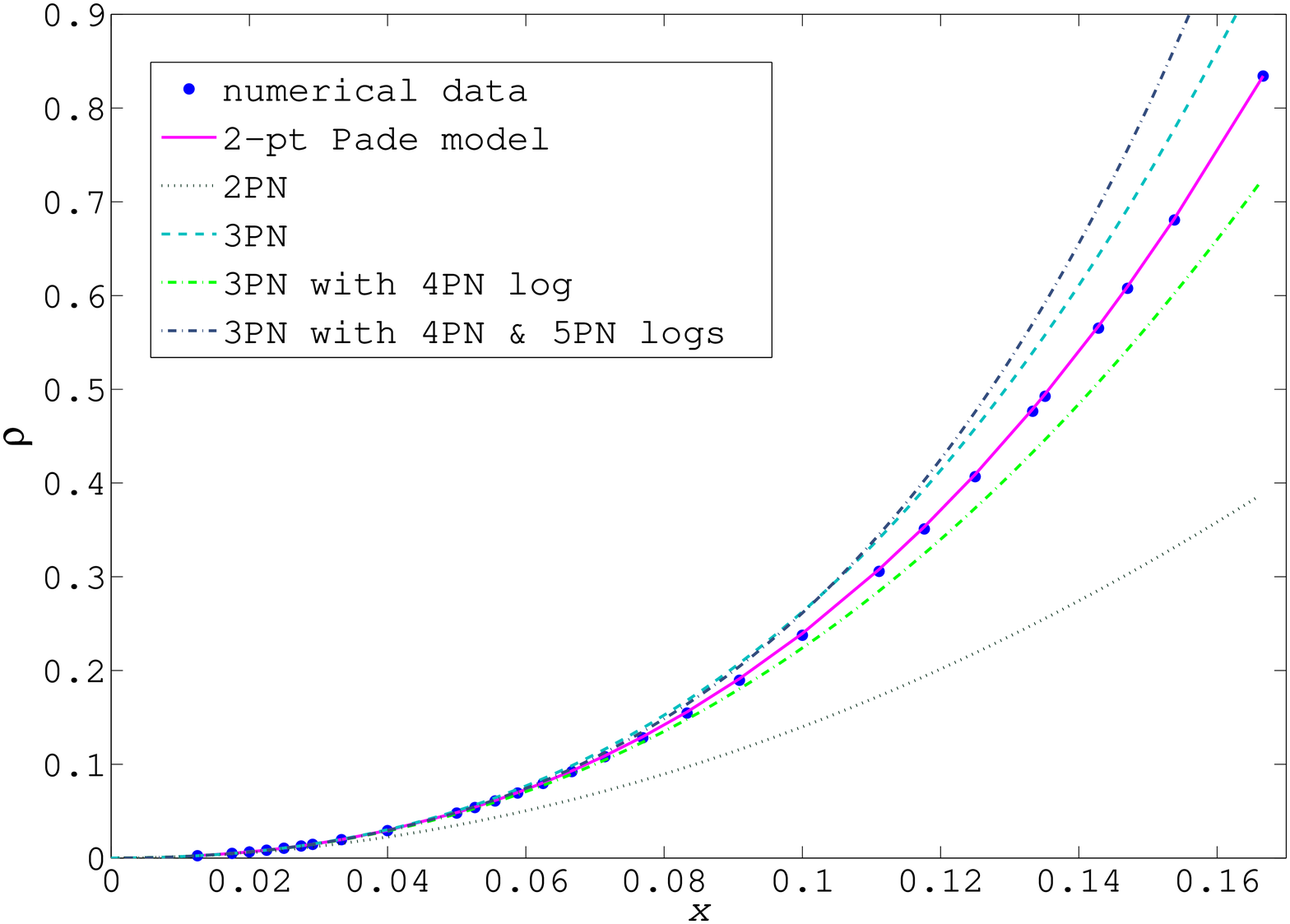}
\caption{Global fit for $\rho(x)$ based on the simple 2-point Pad\'e model (\ref{pade2pt}) [with Eq.\ (\ref{2ptvalues})]. Thick (blue) points are the numerical GSF data, while the solid (magenta) line shows the 2-point Pad\'e model, which is based solely on four pieces of information: two at $x=0$ (the 2PN and 3PN coefficients) and two at the ISCO [$\rho(1/6)$ and $\rho'(1/6)$]. For comparison, `broken' curves
show various analytic PN approximations: 2PN (dotted), 3PN (dashed), 3PN including the 4PN logarithmic term (dash-dot, below the numerical data points; light green online) and 3PN including both 4PN and 5PN logarithmic terms (dash-dot, top curve; dark blue online).
Evidently, a mere knowledge of the GSF (and its derivative) at $x=1/6$ already improves significantly our ability to construct a faithful model of $\rho$ in the strong-field regime. 
}
\label{fig:pade}
\end{figure}

Let us start by considering a very simple ``two-point'' Pad\'e approximant of $\rho(x)$, 
based only on readily available $\rho$-information given at $x=0$ and at the ISCO, $x=1/6$. Specifically, let us try to base our approximant on only 4  ``easy'' pieces of information, namely the values $\rho''(0)$, $\rho'''(0)$, $\rho(1/6)$ and $\rho'(1/6)$ (where, recall, a prime denotes $d/dx$).  The first two values [which themselves assume the knowledge that $\rho(0)=0=\rho'(0)$]
are simply related to the 2PN and 3PN parameters $\rho_2$ and $\rho_3$, which are given analytically in Eq.\ (\ref{rho23}). The latter two values are obtained from the strong-field GSF data: $\rho(1/6)$ is given in the last line of Table \ref{Table:data}, and to obtain the derivative $\rho'(1/6)$ we used a quadratic fit based on 18 near-ISCO data points between $x=1/6.05$ and $x=1/6.0005$ [the same data points used to obtain $\rho(1/6)$ itself]. We find $\rho'(1/6)\simeq 12.66$. We then consider a simple 4-parameter Pad\'e model of the form
\begin{equation} \label{pade2pt}
\rho(x)=c_0x^2\left(\frac{1+c_1x}{1+d_1x+d_2x^2}\right),
\end{equation}
where we have factorized the leading-order PN power $x^2$. The 4 model parameters $\{c_0,c_1,d_1,d_2\}$ are uniquely determined by our 4 ``data points''. Their values are found to be
\begin{equation}\label{2ptvalues}
c_0=\rho_2=14, \quad\quad
c_1= 13.3687, \quad\quad
d_1= 4.60958,\quad\quad
d_2= - 9.47696.
\end{equation}
The resulting 2-point-based Pad\'e model for $\rho(x)$ is plotted in Figure \ref{fig:pade} against the actual numerical data. The agreement appears strikingly good throughout the entire $x$ range, despite the fact that our model only uses information at the two ``end'' points $x=0$ and $x=1/6$. The $L^{\infty}$-norm of the full numerical GSF data, with respect to this model, is found to be as small as $\sim 2.4\times 10^{-3}$. Note that this is within the error tolerance stated above, of $<2.8\times 10^{-3}$. It is remarkable that our extremely simple model (\ref{pade2pt})---which is based merely on 4 readily available pieces of PN and GSF information---would appear to be perfectly adequate for our current purpose (at least in the interval $0\leq x\leq 1/6$). 

In the above analysis we may, of course, replace the derivative $\rho'(1/6)$ with the value of $\rho$ at a different strong-field point---say $x=1/8$---which can give us a better handle on the ``curvature'' of $\rho(x)$.  The quartet $\{\rho_2,\rho_3,\rho(1/8),\rho(1/6)\}$ again determines the parameters $\{c_0,c_1,d_1,d_2\}$ in the model (\ref{pade2pt}) uniquely. This time we find $L^{\infty}\sim 2.4\times 10^{-4}$, representing a full order-of-magnitude improvement with respect to the 2-point model considered before. Such a 3-point Pad\'e model, which still uses a minimal amount of readily-available GSF data, now qualifies for the development of EOB models which are ten times as accurate (this might become necessary in the future). We may obtain similar 3-point models replacing $\rho(1/8)$ with other strong-field data points, and in Table \ref{Table:Pade} we list a few such models, showing in each case the values of the model parameters as well as the corresponding $L^{\infty}$-norm (over the
interval $0< x\leq 1/6$). Such models yield similar values of $L^{\infty}$-norm (with the best performance apparently achieved taking the third data point to be in the vicinity of $x=1/9$). To improve the model further requires additional parameters and, commensurably, additional data points. The last two lines of Table \ref{Table:Pade} present examples of 4- and 6-point models. 
\begin{table}[htb]
\begin{tabular}{l|cccccc|l}
\hline\hline
data points used  & $c_1$  & $c_2$ & $c_3$ & $d_1$ & $d_2$ & $d_3$ & $L^{\infty}$-norm \\
\hline\hline
$x=\left\{0,\frac{1}{6}\right\}$    &
$13.3687$ & -- & -- & $4.60958$ & $-9.47696$ & -- & $2.4\times 10^{-3}$ \\
$\left\{0,\frac{1}{7},\frac{1}{6}\right\}$    &
$14.4329$ & -- & -- & $5.67378$ & $- 12.8852$ & -- & $2.4\times 10^{-4}$ \\
$\left\{0,\frac{1}{8},\frac{1}{6}\right\}$    &
$ 14.4455$ & -- & -- & $5.68636$ & $- 12.9256$ & -- & $ 2.3\times 10^{-4}$ \\
$\left\{0,\frac{1}{9},\frac{1}{6}\right\}$    &
$  14.4892$ & -- & -- & $5.73013$ & $- 13.0657$ & -- & $ 2.0\times 10^{-4}$ \\
$\left\{0,\frac{1}{10},\frac{1}{6}\right\}$    &
$  14.5502$ & -- & -- & $5.79112$ & $- 13.2611$ & -- & $ 2.1\times 10^{-4}$ \\
$\left\{0,\frac{1}{11},\frac{1}{6}\right\}$    &
$  14.6103$ & -- & -- & $5.85123$ & $- 13.4536$ & -- & $ 3.4\times 10^{-4}$ \\
$\left\{0,\frac{1}{12},\frac{1}{6}\right\}$    &
$  14.6726$ & -- & -- & $5.91347$ & $- 13.6529$ & -- & $ 4.6\times 10^{-4}$ \\
$\left\{0,\frac{1}{10},\frac{1}{8},\frac{1}{6}\right\}$    &
$  16.1885$ & $12.0604$ & -- & $7.42943$ & $- 12.8852$ & -- & $ 1.3\times 10^{-4}$ \\
$\left\{0,\frac{1}{14},\frac{1}{12},\frac{1}{10},\frac{1}{8},\frac{1}{6}\right\}$    &
$138.334$ & $1434.67$ & $-866.023$ & $129.575$ & $393.475$ & $-1209.58$ & $ 8.5\times 10^{-5}$ \\
\hline\hline
\end{tabular}
\caption{Simple multiple-point Pad\'e models for $\rho(x)$. Each line of the table corresponds to a particular model, based only on $\rho$-information at the $x$-values indicated in the first column. In all models we have used the known values of the 2PN and 3PN coefficients (the data point ``$x=0$''). For all data points $x\ne 0$ we have used only the value $\rho(x)$ [e.g., for $x=1/7$ we have used $\rho(1/7)$], except in the 2-point model presented in the first line, where we have used the derivative $\rho'(1/6)$ in addition to $\rho(1/6)$. The Pad\'e models have the general form $\rho(x)=14x^2(1+c_1x+c_2x^2+c_3x^3)/(1+d_1x+d_2x^2+d_3x^3)$, where for the 4-point model we set $c_3=d_3=0$ and for the 2,3-point models we additionally set $c_2=0$. For each model, the values of all non-zero parameters $c_n$ and $d_n$ (shown in the table) are uniquely determined by the $\rho$-information assumed.
The last column shows the $L^{\infty}$-norm (in the
interval $0< x\leq 1/6$) of the full numerical GSF data with respect to each of the models. 
}
\label{Table:Pade}
\end{table}

The above experiments illustrate an important point: It is possible to obtain an accurate global model for $\rho(x)$ using the available PN expression in conjunction with just a handful of near-ISCO GSF data points; a Pad\'e approximant based solely on this information can faithfully ``bridge'' over the entire $x$ domain where stable circular orbits exist. This may hold true for other quantities that characterize the $O(\nu)$ dynamics, in which case one has a (relatively) computationally cheap way of modelling such quantities, which does not require an exhaustive GSF survey of the $x$ domain. Of course, in each case one must have at hand some GSF data at intermediate radii in order to be able to confidently assess the faithfulness of one's Pad\'e model. The following general scheme suggests itself: (1) Construct a Pad\'e model (or a set of models) based on all available PN information and on a handful of readily computable strong-field GSF data points; (2) Compute the GSF for a small, complementary sample of intermediate $x$ values and use these data points to assess the performance of the Pad\'e model(s). This scheme could save the need to obtain a dense sample of GSF data across the entire $x$ domain (as we have done in Sec.\ III of this work). Furthermore, in this procedure the ``verification'' GSF data (which may be most computationally expensive) need not be very accurate as in our present work---their accuracy should only match one's error tolerance.

The fits we just presented for the function $\rho(x)$ were mainly given to illustrate a
general strategy for an efficient synergy between GSF, PN and EOB theories. 
In Ref.\ \cite{Damour:2009sm} one of us proposed several other gauge-invariant quantities that may be accessible to both EOB/PN and GSF treatments and that could be used to further constrain the EOB functions at $O(\nu)$. In particular, it was proposed to utilize two gauge-invariant quantities defined for a {\em marginally-bound ``zoom-whirl'' orbit}, namely the GSF-corrected values of the azimuthal ``whirl'' frequency and of the total (dimensionless) orbital angular momentum. It was shown that these two quantities uniquely determine the values of the crucial $O(\nu)$
EOB radial potential $a(x)$, and of its $x$-derivative, at the strong-field point $x=1/4$, i.e.,
significantly beyond the ISCO. This would allow us to extend our strong-field knowledge beyond
the interval $0\leq x\leq 1/6$ that our present study has been limited to. Most importantly, the
application of our general strategy to the function $a(x)$ might give us a good handle on the shape of $a(x)$ in the very-strong-field regime. Indeed, by combining  the current PN knowledge of $a(x)$ near the weak-field point $x=0$ (i.e., the information that $a(0)=a'(0)=a''(0)=0$, together
with the non-zero values of the two numbers $a_3$ and $a_4$) with the hopefully obtainable 
GSF knowledge of  $a(1/4)$ and $a'(1/4)$, we might obtain 
 an accurate global representation of the function $a(x)$ in the entire interval $0\leq x\leq 1/4$.
This would be a most valuable information for the EOB formalism. Then, combining this
global knowledge of the function $a(x)$ with the direct GSF knowledge of the function
 $\rho(x)$ obtained here, we could further derive the global shape of the function
$\bar d(x)$ (at least in the interval $0\leq x\leq 1/6$).

\section{Summary and outlook}

We presented here a calculation of the GSF correction to the precession rate of the orbital periastron for a particle of mass $m$ in a slightly eccentric orbit around a Schwarzschild black hole
of mass ${\mathsf M} \gg m$. Our calculation is {\em exact} at first order in the mass ratio  $q \equiv m/{\mathsf M}$ (within a controlled numerical error of $\lesssim 10^{-4}$ fractionally). Our first main result was the
computation of a dimensionless measure of the $O(q)$ contribution to periastron advance, $\rho$, as
a function of the dimensionless azimuthal-frequency parameter $x=[Gc^{-3}({\mathsf M}+m)\hat\Omega_{\varphi}]^{2/3}$. We computed the function $\rho(x)$ for a dense sample of $x$-values in the interval $0<x\leq 1/6$ corresponding to a sequence of stable circular orbits (down to the ISCO, located at $x=1/6$).

The bulk of this paper concerned itself with a comparison of the GSF results with PN predictions formulated via the EOB formalism, and we explored what can be learned from such a comparison. We demonstrated three main goals that can be achieved by combining GSF and EOB/PN results: (i) test the GSF computation and confirm the EOB/PN results (and, indirectly, also reaffirm the regularization procedures underpinning both GSF and PN theories); (ii) calibrate yet-unknown high-order parameters in the EOB/PN expansion; and (iii) engineer a faithful global model for the dynamical quantity in question, which is valid both in the weak field and in the strong field, and anywhere in between.     

The new quantitative results of this work are (i) the $O(q)$ periastron-advance function $\rho(x)$ (tabulated in Table \ref{Table:data} and plotted in Figure \ref{fig:rho}), (ii) an estimate of the $O(q)$ parts of a few unknown PN parameters [summarized in Eq.\ (\ref{calibrated})], and (iii) an approximate analytic formula for $\rho(x)$ [Given in Eq.\ (\ref{pade2pt}) with Eq.\ (\ref{2ptvalues}); alternative, more accurate, analytic models for $\rho(x)$ are described in Table \ref{Table:Pade}]. 
In addition, we have provided numerical confirmations of several analytically determined PN/EOB
parameters, namely the following  PN expansion coefficients of $\rho(x)$: the 2PN
coefficient $\rho_2$, the 3PN coefficient $\rho_3$, and the recently computed logarithmic
4PN and 5PN coefficients $\rho_4^{\rm log}$ and $ \rho_5^{\rm log}$ \cite{DamourLogs}, whose knowledge
was quite crucial for determining the 4PN and 5PN non-logarithmic contributions.

Our best estimates for the unknown (non-logarithmic) 4PN and 5PN parameters (let alone the 6PN parameter) are rather crude---see the large error bars in Eq.\ (\ref{calibrated}). At fault are two main factors. First, the construction of $\rho(x)$ from the various GSF components, prescribed in Eq.\ (\ref{rhoSFPN}), involves a delicate cancellation of the two dominant terms in the $x$ expansion around $x=0$, namely the $O(x^0)$ term and the $O(x)$ term, leaving a final $\rho(x)$ of $O(x^2)$; we found analytically that  $\tilde F^r_{\rm circ}$ and $\tilde F^r_1$, which are both $\propto x^2$, cancel in the leading  $O(x^0)$ term, while the cancellations occurring in the next (1PN) $O(x)$ term
involve $\tilde F_\varphi^1\propto x^{1/2}$ together with the sub-leading, 1PN corrections to $\tilde 
F^r_{\rm circ}$ and $\tilde F^r_1$.  These cancellations effectively amplify (in relative terms) the numerical error in the small-$x$ GSF data by 2--3 orders of magnitude. The upshot is a rather poor accuracy in $\rho(x)$ for small $x$ values, despite the rather high accuracy of the GSF data itself. The second limiting factor is the restricted capacity of our GSF code to deal with relatively small-$x$ (i.e., large radii) orbits. In our current time-domain architecture, the computational cost increases rapidly with the orbital radius, as a result of the longer evolution time necessary to eliminate the effect of spurious initial radiation. We are currently practically unable to compute the GSF for orbits with radii larger than $\sim 100{\mathsf M}$. Future advances in GSF computational technology (e.g., better treatment of initial conditions, mesh-refinement techniques, or spectral treatments of the perturbation equations---all of which are presently being studied) are certain to reduce the computational cost of GSF calculations, and to allow us to obtain more accurate GSF data and for larger radii. It would be interesting to revisit our analysis once more accurate GSF results are at hand. 

Most importantly, our determination of the {\em function} $\rho(x)$ on the interval $0\leq x\leq 1/6$ has allowed us to compute for the first time the strong-field behavior of a combination of the EOB functions $a(u)$ and $\bar d(u)$, thereby giving us some information of direct significance for constructing accurate
analytical models of the late stages of the dynamics of binary systems.
We have also indicated at the end of the last section how the GSF computation of the
special {\em marginally-bound ``zoom-whirl'' orbit} suggested in  Ref.\ \cite{Damour:2009sm}
might, thanks to the multiple-point Pad\'e strategy introduced here, be used for deriving
the global shape of the crucial $O(\nu)$ radial potential $a(x)$ in the entire domain
$0\leq x\leq 1/4$. Then, combining this
global knowledge of the function $a(x)$ with the direct GSF knowledge of the function
 $\rho(x)$ obtained here, we could further derive the global shape of the function
$\bar d(x)$. This would be a most valuable information for the EOB formalism, and could
significantly help the development of AR models of coalescing binaries.
Unfortunately, our current GSF code cannot handle orbits which are not strictly bound, and we are therefore unable to supply the necessary GSF data at the present time. We are, however, investigating ways to make this calculation possible, and we are hoping to present such analysis in a future paper.

Finally, we recall that we now have at hand a GSF code for orbits with arbitrary eccentricities \cite{Barack:2010tm}. In a forthcoming paper \cite{BSprep}, two of us will present a computation of a certain gauge-invariant relation characterizing the conservative $O(q)$ dynamics of such orbits. A GSF--EOB analysis may then, for the first time, allow us access to the strong-field behavior
of the EOB $Q$ function, which enters the description of the ``radial'' component of the binary motion. It is our impression that there are profound prospects for further fruitful synergy between the GSF and EOB frameworks, and we expect this activity to gain considerable momentum in the coming years.

\section*{ACKNOWLEDGEMENTS}
LB acknowledges support from STFC through Grant No.~PP/E001025/1, and wishes to thank the Institut des Hautes \'Etudes Scientifiques, where part of the work on this project took place, for hospitality and financial support. NS acknowledges support from Monbukagaku-sho Grant-in-Aid for the global COE program {\it The next generation of physics, spun from universality and emergence}. 



\end{document}